\pdfoutput=1 
\documentclass[aps,prd,twocolumn,superscriptaddress,amsfont,graphicx,nofootinbib,preprintnumbers]{revtex4}

\usepackage{amsmath}
\usepackage{amssymb}
\usepackage{color}
\usepackage{xcolor}
\usepackage{graphicx}
\usepackage{hyperref}
\usepackage{textcomp}
\usepackage{xspace,slashed}
\usepackage[utf8]{inputenc}
\usepackage{subcaption}
\usepackage{multirow}
\usepackage{nccmath} % \mfrac
\usepackage{slashed}
\usepackage[bb=boondox]{mathalfa} % \mathbb{1}
\usepackage{orcidlink}

\newcommand{\ep}{\epsilon}
\newcommand{\tpi}{2 \pi i \,}
\newcommand{\G}{\mathcal{G}}
\newcommand{\B}{\mathcal{B}}
\newcommand{\kallen}{\lambda}
\newcommand{\bc}{f}

\begin{document}

\title{One-loop Feynman integrals from their maximal cut using dispersion relations}

\author{Mohammed Alsadah}
\email{mohammedalsadah31415@gmail.com}
\affiliation{School of Physics and Astronomy, The University of Edinburgh, Edinburgh EH9 3FD, Scotland, UK}
\author{Piotr Bargie\l{}a~\orcidlink{0000-0002-3646-5892}}
\email{piotr.bargiela@stonybrook.edu}
\affiliation{C.N. Yang Institute for Theoretical Physics, Stony Brook University, Stony Brook NY 11794-3840 USA}
\affiliation{Higgs Centre for Theoretical Physics, School of Physics and Astronomy, The University of Edinburgh, Edinburgh EH9 3FD, Scotland, UK}
\author{Miko\l{}aj Marci\'{o}w~\orcidlink{0009-0007-5299-3330}}
\email{mikolaj.marciow@pwr.edu.pl}
\affiliation{Wroc\l{}aw University of Science and Technology, Wyb. Wyspia\'{n}skiego 27, 50-370 Wroc\l{}aw, Poland}
\affiliation{Max Planck Institute for the Physics of Complex Systems, 
01187 Dresden, 
Germany}

\begin{abstract}

In this work, we present a new dispersive formula that retrieves from its maximal cut any multi-point one-loop scalar Feynman integral in the non-anomalous region with unit powers of standard squared propagators and unit numerator.
It requires performing less integrations of an algebraic integrand than other representations and it does not require performing phase space integrals.
We write explicit realization for all examples relevant for collider phenomenology, including the new triple dispersion relation for pentagon and hexagon integrals.
We also find and analyse a new anomalous region for the triangle integral.

\end{abstract}

\maketitle 

\allowdisplaybreaks

\tableofcontents

%======================================================================
\section{Introduction}

Feynman integrals are an inherent part of high-precision theoretical predictions in perturbative Quantum Field Theory.
We focus here on Standard Model integrals relevant for collider phenomenology.
For a review of recent major progress in integral properties and results, see Refs~\cite{Heinrich:2020ybq, Bourjaily:2022bwx}.
While multi-loop integrals are required to achieve state-of-the-art accuracy, their general analytic structure has not been fully unveiled yet.
For this reason, the further analysis of one-loop integrals is considered an important starting point.

The analytic properties of one-loop Feynman integrals are well-understood up to the finite part in dimensional regulator $\ep=(d_0-d)/2$ in $d_0=4$.
Analytically, all divergent integrals are provided by \texttt{QCDloop}~\cite{Ellis:2007qk}, while the finite ones can be calculated using \texttt{Package-X}~\cite{Patel:2016fam}.
Beyond finite part, one-loop integrals can be computed analytically with e.g. \texttt{HyperInt}~\cite{Panzer:2014caa} or \texttt{SubTropica}~\cite{Giroux:2026tgd}, or numerically using e.g. \texttt{AMFlow}~\cite{Liu:2022chg}.
Using Passarino-Veltman~\cite{Passarino:1978jh} or Integration-By-Parts~\cite{Tkachov:1981wb, Chetyrkin:1981qh} reductions, one can linearly relate all one-loop integrals to only those with unit propagator powers and unit numerator.
Moreover, after an appropriate integrand-level decomposition in four dimensional theory, only the integrals with up to four propagators contribute to the finite part of any one-loop scattering amplitude~\cite{Bern:1992em}, and only those with up to five propagators contribute beyond~\cite{Bern:1993kr, Bargiela:2024rul}.
This leaves only five independent functions, i.e. tadpole, bubble, triangle, box, and pentagon, up to kinematic degeneracies, and we consider all of them in this work.

A particularly suitable way to study the analytic properties of Feynman integrals is through their dispersive representation~\cite{Mandelstam:1958xc, mandelstam1959construction, Mandelstam:1959bc}.
Indeed, it manifests branch cut structure in the integration boundary, as well as singularities of the discontinuity in the integrand, see example variety of modern applications in Refs~\cite{Pire:2010if, Colangelo:2015ama, remiddi2016differential, abreu2017cuts, Homrich:2019cbt, Correia:2020xtr, Correia:2022dcu, Vergu:2023rqz, Tourkine:2023xtu, Dawid:2024oey, Hoferichter:2024vbu, Herren:2026xfq}.
In practice, the integration region and the integrand are nontrivial to find because it requires performing the phase space integration over the on-shell cut propagators.
However, both of these difficulties can be avoided if the integrand is the maximal cut of the integral, which is algebraic at one loop.
In this work, we present and exemplify the dispersive integration region that retrieves the full uncut integral from its maximal cut.

Understanding the analytic properties of Feynman integrals is vital to multiple evolving research directions.
For example, it allows the Landau bootstrap~\cite{Eden:1966dnq, Hannesdottir:2024hke, McLeod:2026kaf} program to recover the integral using its Landau singularity~\cite{Landau:1959fi, Correia:2025yao} structure.
In addition, it aligns with the revival of the analytic S-matrix program~\cite{Correia:2021etg, Hannesdottir:2022bmo, Caron-Huot:2023ikn}.
Moreover, this work is a major step towards Multivariate Integrated Unitarity, an approach to calculating scattering amplitudes in a new sector-by-sector manner at the integrated level, started in Ref.~\cite{Bargiela:2024hgt}, that builds on the integrand-level Generalized Unitarity~\cite{Britto:2004nc, Anastasiou:2006jv, Forde:2007mi}.

%======================================================================
\section{Dispersive formula}

Consider a scalar one-loop $n$-point Feynman integral with unit powers of standard squared propagators and unit numerator
\begin{equation}
I_n
= \int \frac{d^d k}{(2\pi)^d} \, \frac{1}{(q_1^2-m_1^2) \cdots (q_n^2-m_n^2)} \,,
\end{equation}
where the propagator momentum
$q_j = k + p_1 + \dots + p_{j-1} \,$,
depends on the loop momentum $k$ and the corresponding external momenta $p_i\,$, while $m_j$ is a mass of the propagator $j$.
The integral $I_n$ in $d_0=4$ is a function of $n$ masses $m_j^2$ and at most $4n-10$ Mandelstam invariants e.g. $s_{ij}=p_{ij}^2\,$, $s_{ijk}=p_{ijk}^2\,$, including the $n$ off-shell variables $s_{ii}=p_{ii}^2=4p_i^2\,$, where we define $p_{ij}=p_i+p_j$ and $p_{ijk}=p_i+p_j+p_k$ for convenience.
The function $I_n$ can be written as Cauchy's integral $I_n(x) = \oint_C \frac{d x' I(x')}{\tpi(x'-x)}$ in each kinematic variable $x$.
When the Cauchy contour $C$ is stretched to infinity, the remaining integral is over the discontinuity in the corresponding kinematic variable and along the corresponding singular branch cut region~\footnote{If the integral diverges at infinity, convergence can be restored with the so called subtraction terms.}.
One can iterate this procedure in enough Mandelstam variables so that the integrand becomes the total discontinuity of $I_n$.
This total discontinuity is proportional to the maximal cut of $I_n$~\cite{abreu2017cuts} which is much easier to calculate than the original integral, especially at the one-loop order where it is algebraic and not transcendental.
The maximal cut $M_n$ is defined by putting all the propagators $j$ on shell by following the Cutkosky rule
$(q_j^2-m_j^2)^{-1} \to -\tpi \delta^+(q_j^2-m_j^2) \,$~\cite{Cutkosky:1960sp}.
The resulting well-known formula yields~\cite{Bosma:2017ens, Harley:2017qut}
\begin{equation}
M_n = c'_n \, \G_n^{(n-d)/2} \, \B_n^{(d-n-1)/2} \,,
\end{equation}
with an overall coefficient
$c_n' = \pi^{(1-n)/2} (4\pi)^{-d/2} / \Gamma(d-n+1) \,$,
where
\begin{equation}
\G_n = 
\left| \begin{array}{cccc}
p_1\cdot p_1  & p_1\cdot p_2& \cdots & p_1 \cdot p_{n-1}  \\
p_2\cdot p_1  & p_2\cdot p_2& \cdots & p_2 \cdot p_{n-1} 
\\
\vdots & \vdots &  \ddots & \vdots \\ 
p_{n-1}\cdot p_1  & p_{n-1} \cdot p_2& \cdots & p_{n-1} \cdot p_{n-1} 
\end{array}
\right|
\end{equation}
is the Gram determinant, and
\begin{equation}
\B_n = 
\left| \begin{array}{cccc}
k\cdot k  & k\cdot p_1& \cdots & k \cdot p_{n-1}  \\
p_1\cdot k  & p_1\cdot p_1& \cdots & p_1 \cdot p_{n-1} 
\\
\vdots & \vdots &  \ddots & \vdots \\ 
p_{n-1}\cdot k  & p_{n-1} \cdot p_1& \cdots & p_{n-1} \cdot p_{n-1} 
\end{array}
\right|_{\text{max cut}}
\end{equation}
is the Baikov polynomial~\cite{Baikov:1996iu} evaluated on the maximal cut $q_j^2=m_j^2$ for all propagators $j$.
The Baikov polynomial is a function of propagator masses and Mandelstam invariants, whereas the Gram polynomial is only a function of the latter.
They are both quadratic in each variable.

We present a new dispersive formula that retrieves any integral $I_n$ in the non-anomalous (defined below) region from its maximal cut $M_n$
\begin{equation*}
\begin{split}
\raisebox{-3.6em}{\includegraphics[width=0.15\textwidth]{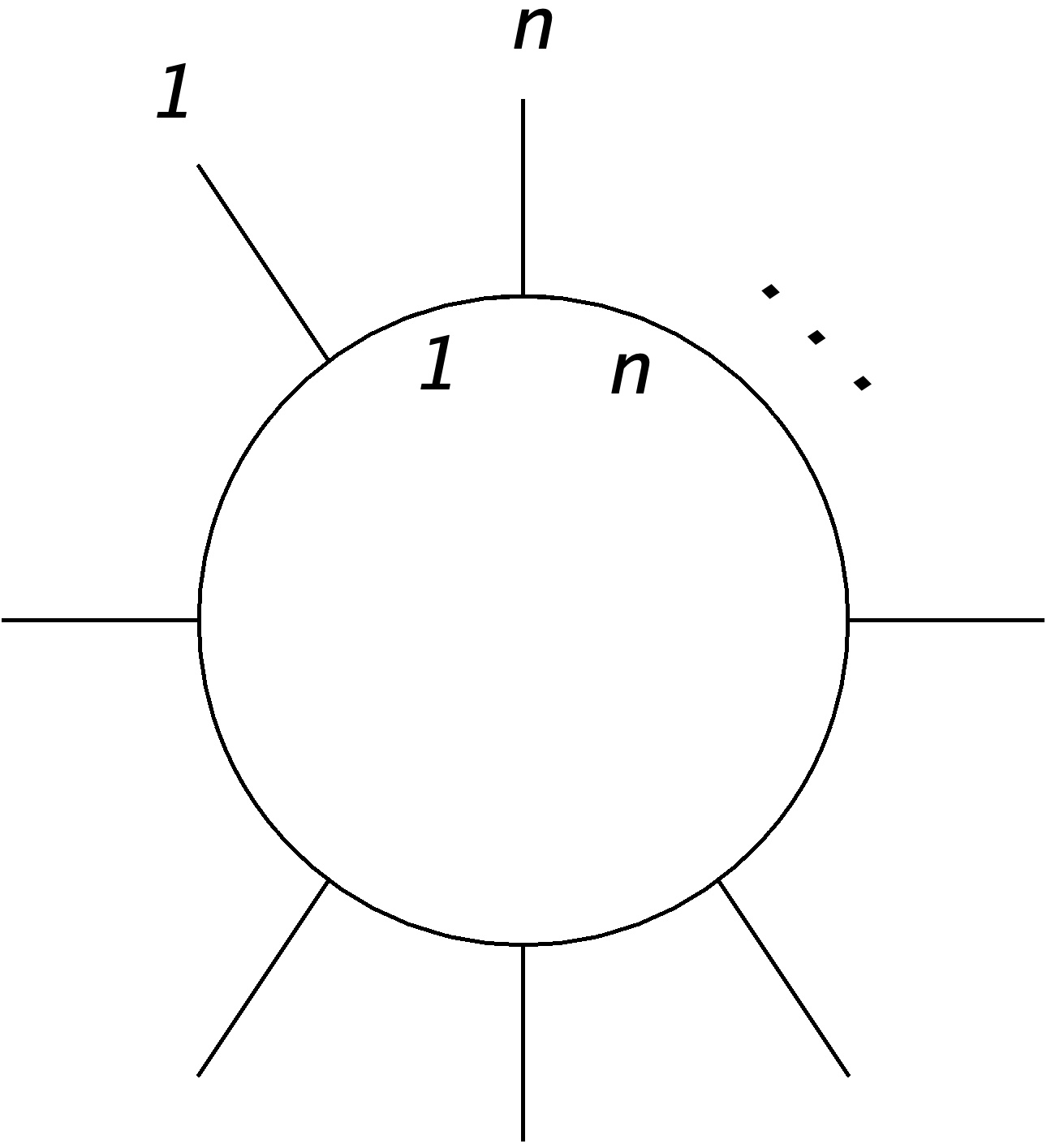}}
= \int
\raisebox{-3.9em}{\includegraphics[width=0.15\textwidth]{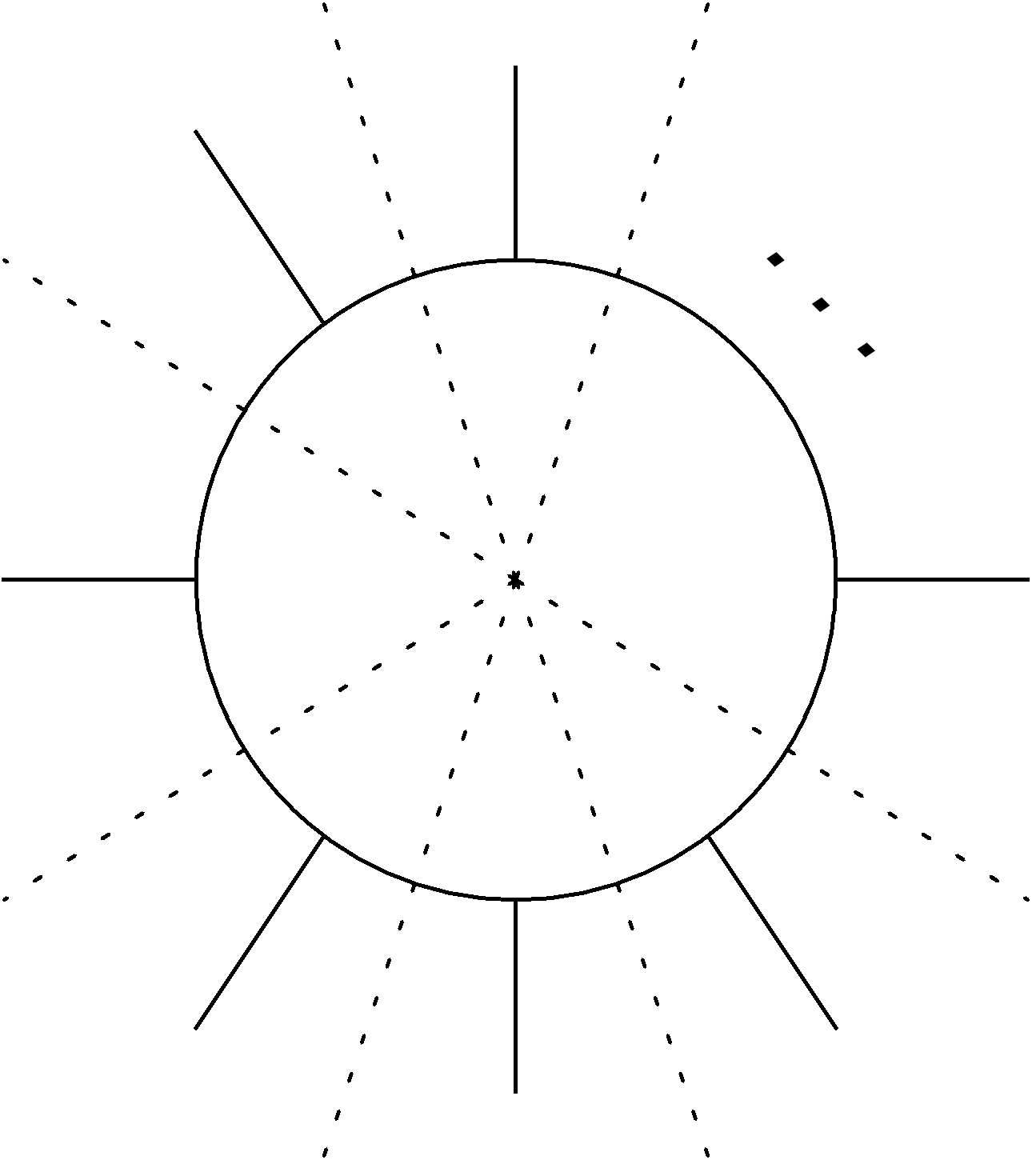}}
\end{split}
\end{equation*}
\begin{equation}
I_n(\vec{x})
= \tilde{c}_n
\int_{(m_0+\tilde{m}_0)^2}^\infty \frac{d x_0'}{x_0'-x_0} 
\int_{B_+} \prod_{i=1}^{N-1} 
\frac{d x_i'}{x_i'-x_i} 
M_n(\vec{x'}) \,,
\label{eq:disp}
\end{equation}
with $N = \lceil n/2 \rceil\,$ cuts, an overall coefficient $\tilde{c}_n$ specified for each example with
$c_n = 2 \pi^n / (2\pi)^N \,$,
where we collectively denote all variables $s_{jk}$ and $m_j^2$ by $\vec{x}\,$, while $\vec{x'}$ differs from $\vec{x}$ in the first $N$ entries that are integration variables $s_{jk}'$ corresponding to the maximal cut, and the most outer integration region in $x_0'$ is given by the unitarity production threshold across the two propagators with masses $m_0$ and $\tilde{m}_0$ that are cut in this channel.
The integration region $B_+$ over the $N-1$ inner integration variables is defined by a real solution to $(-1)^N \B(\vec{x'})>0\,$ laying above the unitarity production threshold in corresponding cut channels.

Let us make a few comments on the formula~\eqref{eq:disp}.
Firstly, it degenerates for $n=1,2$.
Indeed, for $n=1\,$, there is simply no dispersion relation in external kinematics because there is no production channel, while, for $n=2\,$, the most inner integral is not required since there is only one cut.
Secondly, for $n>2\,$, the maximal cut $M_n$ in the integrand can be obtained in principle via multiple different cut channel compositions.
Therefore, one can choose the dispersive channels according to the problem at hand, as long as the combined cuts go through each propagator at least once.
Thirdly, the integration region simplifies to $(0,\infty)$ in each variable for massless propagators at $n=2,3\,$, and also with on-shell legs for $n>2$.
Fourthly, this representation provides a way to integrate the algebraic integrand the smallest number of times for $n>3\,$, since, in comparison, $n-1$ integrals are required in standard Feynman, while $n$ in Baikov parametrization~\cite{Baikov:1996iu}.
Fifthly, our novelty in the dispersive representation~\eqref{eq:disp} is the explicit algebraic definition of the integration region, contrary to the standard solution to the on-shell cut conditions in the momentum phase space integral, and its particular realization for $n>4$ which was never considered before.
Finally, for $n>2\,$, the formula~\eqref{eq:disp} gets additional contributions from different integration contours when the kinematic point $\vec{x}$ is in an anomalous region.
We expand on it later in this manuscript.

%======================================================================
\section{Examples}

In this section, we write an explicit form of the dispersion relation~\eqref{eq:disp} for all independent one-loop integrals relevant to the collider phenomenology.
Note that the dispersive representation for the considered generic examples holds also for specialized more degenerate kinematics in the corresponding limit.
If not stated otherwise, we have checked the presented dispersive formulae on a generic convergent rational probe in integer dimensions in four cases of either fully massive or massless internal kinematics in propagators or external momenta and found agreement to at least 10 significant digits against a numerical evaluator~\cite{Patel:2016fam, Liu:2022chg}.

\subsection{Single-cut bubble}

Let us start by considering the well-understood 2-point bubble integral $I_2$ with generic nonzero propagator masses $m_1,m_2$ and momentum $p^2$~\cite{davydychev1992general, shpot2007massive, remiddi2016differential, anisovich1993bethe}.
Its maximal cut reads
\begin{equation}
M_2 = c_2' (p^2)^{(2-d)/2} 
\left( -\frac{1}{4} \kallen(p^2,m_1^2,m_2^2) \right)^{(d-3)/2} \,,
\end{equation}
where
\begin{equation}
\begin{split}
\kallen(a^2,b^2,c^2) = (a^2-(b+c)^2) (a^2-(b-c)^2)
\end{split}
\end{equation}
is the K\"{a}ll\'{e}n function, $(m_1+m_2)^2$ is the unitarity production threshold, while $(m_1-m_2)^2$ is the pseudothreshold.
The dispersive representation
\begin{equation}
\small
I_2(p^2,m_1,m_2) 
= i^{d-2} c_2 \int_{(m_1+m_2)^2}^\infty 
\frac{dp'^2}{p'^2-p^2} M_2(p'^2,m_1,m_2)
\end{equation}
retrieves $I_2$ from a single cut $M_2$ through both propagators $q_1,q_2$ integrated along the unitarity production threshold.
In this example, there is only one unique dispersive representation. 
As there are no anomalous thresholds here, this formula holds for all values of the kinematic variables, and its convergent form yields
\begin{equation}
\small
\begin{split}
&I_2 = \frac{2^{1-d} i \pi^{-d/2} \Gamma\left(2-\frac{d}{2}\right) (m_1m_2)^{\frac{d-3}{2}} (m_1+m_2)}{(m_1+m_2)^2-s} \\
&\times F_1\left(\frac{d-1}{2},\frac{3-d}{2},1,\frac{3}{2};-\frac{(m_1-m_2)^2}{4m_1m_2};\frac{s}{s-(m_1+m_2)^2}\right) \,.
\end{split}
\end{equation}

\subsection{Double-cut triangle}

Consider the 3-point triangle integral $I_3$ with generic nonzero propagator masses $m_1,m_2,m_3$ and momenta $p_1^2,p_2^2,p_3^2$~\cite{fronsdal1964integral, ball1991form, Coriano:1994sz, Lucha:2006vc, Suzuki:2008hm}.
For convenience, from now on, we will suppress function arguments that are not important at a given stage.
The dispersive representation
\begin{equation}
I_3
= i^d c_3 \int_{(m_2+m_3)^2}^\infty \frac{dp_2'^2}{p_2'^2-p_2^2} 
\int_{\bc_1^-}^{\bc_1^+} \frac{dp_1'^2}{p_1'^2-p_1^2} M_3
\label{eq:3p}
\end{equation}
retrieves $I_3(p_1^2,p_2^2,p_3^2,m_1,m_2,m_3)$ from a double cut $M_3(p_1'^2,p_2'^2,p_3^2,m_1,m_2,m_3)\,$, i.e. the outer cut through the propagators $q_2,q_3$ integrated along the unitarity production threshold, and the inner cut through the propagators $q_1,q_2$ integrated in the region where $\B_3(p_1'^2) = -m_3^2 (p_1'^2-\bc_{1}^-) (p_1'^2-\bc_{1}^+)/4 > 0$.
As the $\B_3(p_1'^2)$ parabola faces down for all real values of $m_3\,$, we integrate in $p_1'^2$ between the two solutions
\begin{equation}
\small
\begin{split}
\bc_1^\pm(p_2'^2) &= 
\frac{ -\kallen(p_2'^2,m_1^2,m_3^2) -\kallen(p_3^2,m_2^2,m_3^2) + \kallen(p_2'^2,p_3^2,0) }{4 m_3^2} \\
&+ \frac{ \kallen(m_1^2,m_2^2,0) }{4 m_3^2} 
\pm \frac{\sqrt{ \kallen(p_2'^2,m_2^2,m_3^2) \kallen(p_3^2,m_1^2,m_3^2) }}{2 m_3^2} \,.
\end{split}
\end{equation}
Note that this is the simplest example in which the dispersive representation is not unique.
Indeed, one can choose the two integration variables in three different ways, and then choose the integration order in two additional ways, e.g.
\begin{equation}
I_3
= i^d c_3 \int_{(m_1+m_3)^2}^\infty \frac{dp_1'^2}{p_1'^2-p_1^2} 
\int_{\bc_3^-}^{\bc_3^+} \frac{dp_3'^2}{p_3'^2-p_3^2} M_3 \,,
\end{equation}
where $\bc_3^\pm(p_1'^2)$ are now the solutions to $\B_3(p_3'^2)>0\,$.
Moreover, the simplest kinematic configuration for which anomalous thresholds start to appear is the case of two unequal masses among the three nonzero masses~\cite{Lucha:2006vc}.
In addition, the anomalous region depends on the integration variables and order, which may allow avoiding the anomalous contribution by an appropriate representation choice.
We also point out that a single instead of a double dispersion relation is enough to retrieve $I_3$ for some degenerate kinematic cases, e.g. massless one-off-shell triangle, which is proportional to a bubble.
In addition to standard checks, we also found numerical agreement to $\ep^2$ order by expanding around four integer dimensions at $p_3^2=m_1^2-m_2^2$.

\subsection{Double-cut box}

Consider the 4-point box integral $I_4$ with generic nonzero propagator masses $m_1,m_2,m_3,m_4\,$, momenta $p_1^2,p_2^2,p_3^2,p_4^2$ and Mandelstam invariants $s_{12},s_{23}$~\cite{mandelstam1959construction, DeTollis:1964una, js1970analytic, frederiksen1973analytic}.
The dispersive representation reads
\begin{equation}
I_4
= i^{d-3} c_4 \int_{(m_2+m_4)^2}^\infty \frac{ds_{23}'}{s_{23}'-s_{23}} 
\int_{\bc_{12}^+}^\infty \frac{ds_{12}'}{s_{12}'-s_{12}} M_4 \,,
\end{equation}
where the inner integral over the the double cut $M_4$ is along the upper-unbounded disconnected solution to $\B_4(s_{12}') = a_{12} (s_{12}'-\bc_{12}^-) (s_{12}'-\bc_{12}^+) > 0\,$, as $a_{12}>0$ for $s_{23}'>(m_2+m_4)^2\,$, and the lower-unbounded solution does not contribute as it lies below the production threshold $s_{12}'>(m_1+m_3)^2$.
Note that this is also the simplest example in which the dispersive representation is different in Mandelstam variables than in squares of external momenta.
Indeed, as
\begin{equation*}
\begin{split}
\raisebox{-1.7em}{\includegraphics[width=0.15\textwidth]{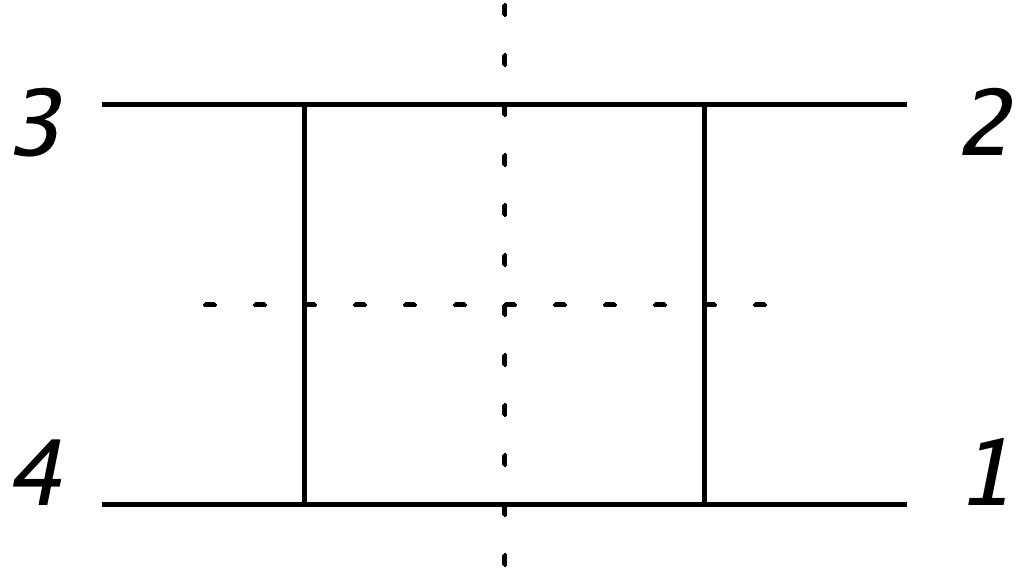}}
= 
\raisebox{-2.0em}{\includegraphics[width=0.15\textwidth]{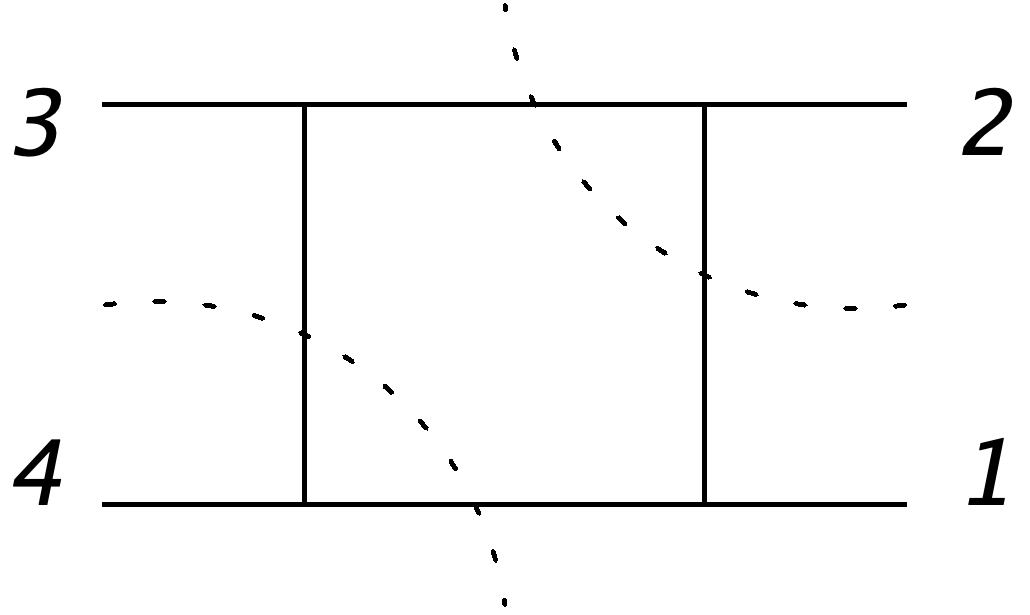}} \,,
\end{split}
\end{equation*}
then we can also write
\begin{equation}
I_4
= i^{d-3} c_4 \int_{(m_3+m_4)^2}^\infty \frac{dp_3'^2}{p_3'^2-p_3^2} 
\int_{\bc_{1}^+}^\infty \frac{dp_1'^2}{p_1'^2-p_1^2} M_4 \,.
\end{equation}
In addition to standard checks, for massless propagators, we also found numerical agreement to $\ep^2$ order and analytic agreement at $\ep^0$~\cite{Duplancic:2002dh, Denner:2010tr}.

\subsection{Triple-cut integrals}

In this subsection, we analyse new triple dispersion relation examples.

Consider the 5-point pentagon integral $I_5$ with generic nonzero propagator masses $m_1,m_2,m_3,m_4,m_5\,$, momenta $p_1^2,p_2^2,p_3^2,p_4^2,p_5^2$ and Mandelstam invariants $s_{12},s_{23},s_{34},s_{45},s_{15}$.
The triple dispersive representation reads
\begin{equation}
\begin{split}
&\raisebox{-1.7em}{\includegraphics[width=0.13\textwidth]{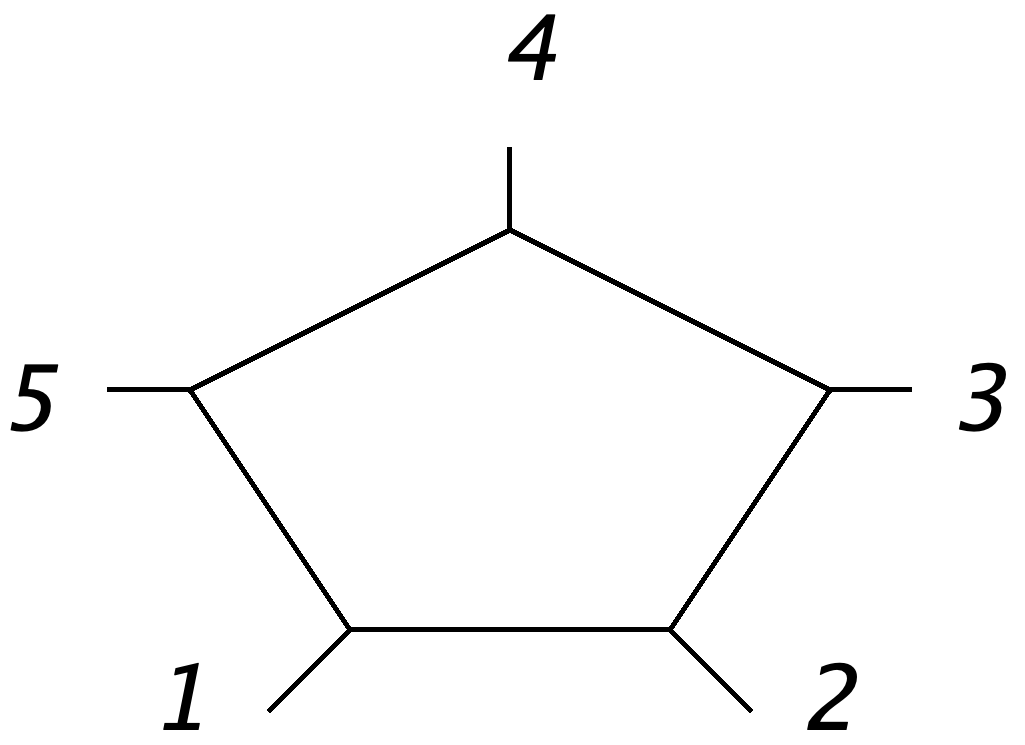}}
= i^{d-3} c_5
 \int_{(m_1+m_3)^2}^{\infty} \frac{ds_{12}'}{s_{12}'-s_{12}} \\
& \times \int_{\bc_{23}^+}^\infty \frac{ds_{23}'}{s_{23}'-s_{23}} 
\int_{\bc_{15}^-}^{\bc_{15}^+} \frac{ds_{15}'}{s_{15}'-s_{15}}
\raisebox{-1.9em}{\includegraphics[width=0.13\textwidth]{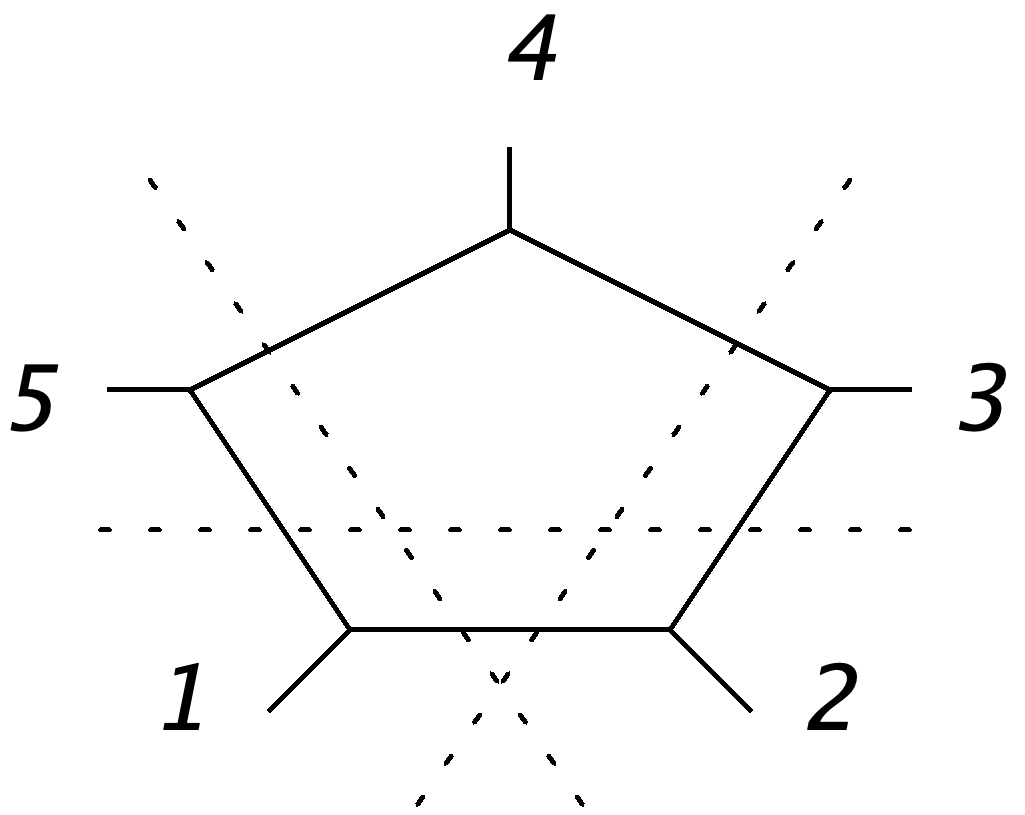}} \,,
\end{split}
\end{equation}
where the most inner integral between $\bc_{15}^\pm(s_{12}',s_{23}')$ over the triple cut $M_5$ is along the solution to $\B_5(s_{15}') = a_{15} (s_{15}'-\bc_{15}^-) (s_{15}'-\bc_{15}^+) < 0\,$ for $a_{15}>0\,$, and the middle integral is along $s_{12}' \in (\bc_{23}^+(s_{12}') , \infty)\,$, i.e. the upper-unbounded solution to the reality requirement on $\bc_{15}^\pm(s_{12}',s_{23}')$ formulated through the discriminant condition $\Delta(\B_5(s_{15}'))(s_{23}') = a_{23} (s_{23}'-\bc_{23}^-) (s_{23}'-\bc_{23}^+) > 0\,$ for $a_{23}>0$.
Note that $I_5$ develops an imaginary part in the $s_{23}' \in ((m_2+m_4)^2,\bc_{23}^+)$ integration region.
Due to a poor numerical convergence of the triple dispersive integral, in the most generic fully massive propagator and external momenta case, we found agreement only to 3 significant digits.

A new triple dispersion relation can also be written for the 6-point hexagon integral $I_6\,$, which is relevant e.g. to the three-body scattering~\cite{Dawid:2024oey,Zhang:2026eez}.
For fully massless propagators $m_{1 \dots 6}=0$ and external momenta $p_{1 \dots 6}^2=0\,$, it is a function of Mandelstam invariants $s_{12},s_{23},s_{34},s_{45},s_{56}, s_{16},s_{123},s_{234},s_{345}\,$, and its dispersive representation reads
\begin{equation}
\begin{split}
& \raisebox{-2.3em}{\includegraphics[width=0.15\textwidth]{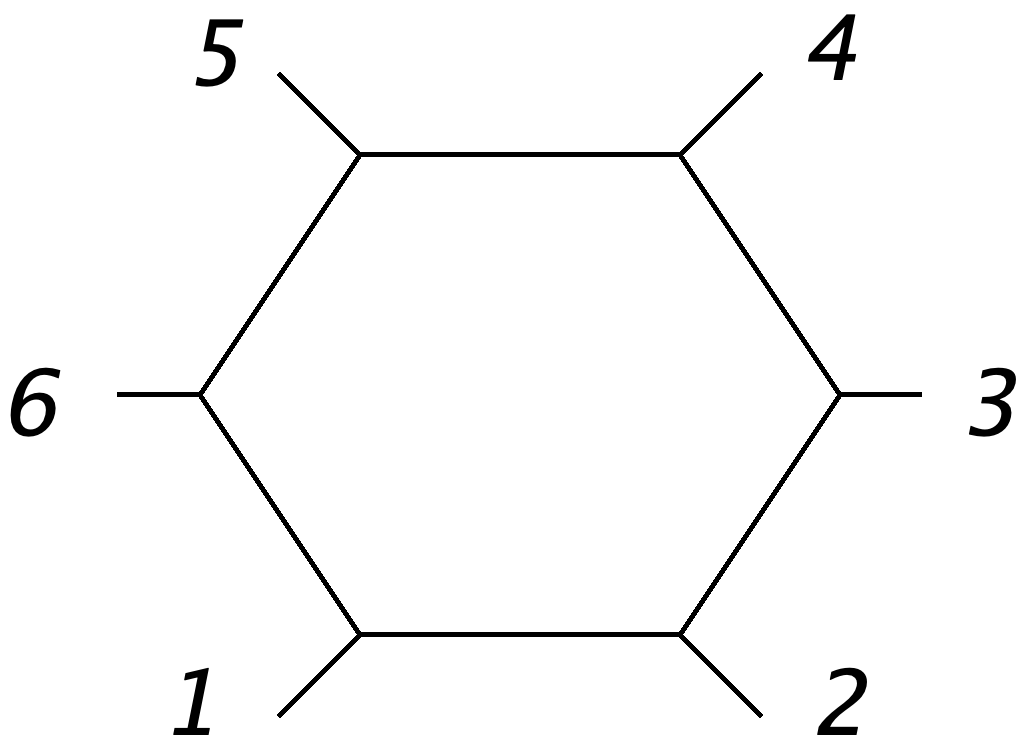}} \,
= \frac{i^{d-3} c_6}{2} 
\int_{0}^{\infty} \frac{ds_{12}'}{s_{12}'-s_{12}} \\
& \times \int_{0}^\infty \frac{ds_{34}'}{s_{34}'-s_{34}} 
\int_{0}^\infty \frac{ds_{56}'}{s_{56}'-s_{56}} \,\,
\raisebox{-2.4em}{\includegraphics[width=0.15\textwidth]{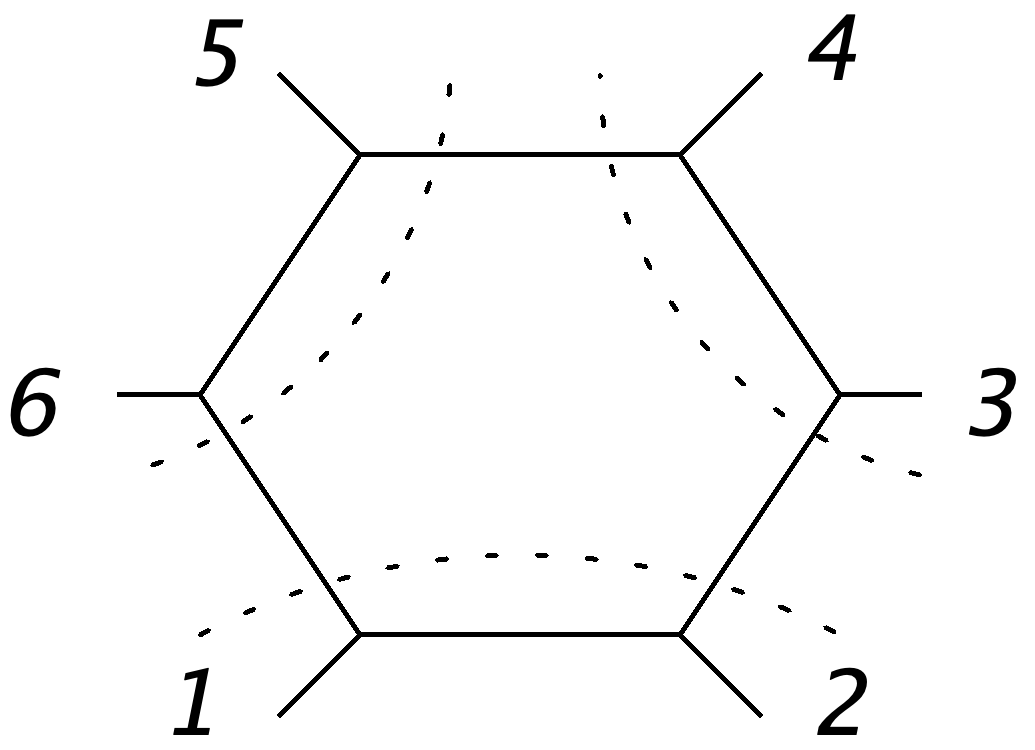}} \,.
\end{split}
\end{equation}

%======================================================================
\section{Anomalous contributions}

Let us elaborate on the region of validity of our general dispersion relation~\eqref{eq:disp}.
If Baikov polynomial singularities for a given choice and order of dispersive integration variables have a nonzero imaginary part, then they are on the secondary Riemann sheet, and thus there is only non-anomalous contribution $I_n$ to the dispersion relation, as in eq.~\eqref{eq:disp}.
In contrast, if Baikov singularities have no imaginary part, then they are on the principal sheet, and thus there is an additional, so called anomalous, contribution $I_{na}$ to the total dispersive formula $I_n + I_{na}$~\cite{PhysRev.111.1187, Mandelstam:1960zz, RevModPhys.33.448, 10.1063/1.1703852, 10.1063/1.1704053, js1970analytic, Goria:2008ny, Passarino:2018wix, Kermanschah:2021wbk}.
In principle, $I_{na}$ can be obtained from $I_n$ by careful analytic continuation.
In practice, finding all the anomalous regions analytically is a complicated task.
Here, we perform the analysis on the $n=3$ triangle integral example.

It is convenient to study the reality of Baikov singularities by the discriminant criterion, i.e. $\Delta(\B(p_1'^2))<0$ for non-anomalous and $\Delta(\B(p_1'^2))>0$ for the anomalous region.
In case of the triangle~\eqref{eq:3p}, we have
\begin{equation}
\Delta(\B(p_1'^2)) = \frac{1}{16} 
\kallen(p_2'^2,m_2^2,m_3^2) \kallen(p_3^2,m_1^2,m_3^2) \,,
\end{equation}
where $\kallen(p_2'^2,m_2^2,m_3^2) \geq 0$ in the whole integration region $p_2'^2 > (m_2+m_3)^2 \geq 0\,$, and thus we find four separate regions, two non-anomalous and two anomalous, by studying the sign of $\kallen(p_3^2,m_1^2,m_3^2)$ i.e.
\begin{equation}
\begin{split}
p_3^2 \in (-\infty,0) \cup ((m_2-m_3)^2,(m_2+m_3)^2) \,, &\text{ nonanom} \,, \\
p_3^2 \in \,\,\,\, (0,(m_2-m_3)^2) \cup ((m_2+m_3)^2,\infty) \,, &\text{ anom} \,,
\end{split}
\end{equation}
where the last anomalous region is our new finding, which we describe and check below.

Let us analyse the new anomalous region $p_3^2 > (m_2+m_3)^2$ in $d=4$.
The corresponding anomalous contribution to the dispersive representation yields 
\begin{equation}
\small
I_{3a} = 
2 \, c_3 \,  \int_{f_2^*}^\infty \frac{dp_2'^2}{p_2'^2-p_2^2} \mathcal{P}
\int_{\bc_1^+}^{\tilde{\bc}_1^-} \frac{dp_1'^2}{p_1'^2-p_1^2} M_3
\end{equation}
where the maximal cut reads $M_3 = c'_3 (\sqrt{\G_3})^{-1}\,$, the overall factor of 2 results from taking the discontinuity of the power function in the integrand, the new upper integration boundary of the inner integral is the smaller of the two singular points 
$\tilde{\bc}_1^\pm = (\sqrt{p_2'^2} \pm \sqrt{p_3^2})^2$ 
of the Gram polynomial
$\G(p_1'^2) 
= -\kallen(p_1'^2,p_2'^2,p_3^2)/4
= -(p_1'^2 - \tilde{\bc}_1^-) (p_1'^2 - \tilde{\bc}_1^+)/4 \,$
in the integrand, while the new lower integration  boundary of the outer integral
\begin{equation}
\begin{split}
f_2^* = 
&\frac{ \kallen(p_3^2,m_1^2,m_3^2)+ 2 p_3^2 (m_2^2+m_3^2) }{2 p_3^2} \\
&+\frac{(p_3^2 - m_1^2 + m_3^2) \sqrt{  \kallen(p_3^2,m_1^2,m_3^2) + 4 p_3^2 m_2^2 }}{2 p_3^2}
\end{split}
\end{equation}
is defined by the larger of the two solutions to
$(\tilde{\bc}_1^- - \bc_1^+)|_{p_2'^2=f_2^*} = 0 \,$.
Since $\bc_1^- \leq \bc_1^+ \leq \tilde{\bc}_1^- < \tilde{\bc}_1^+$ for $p_2'^2>(m_2+m_3)^2\,$, the anomalous Gram integrand singularities $\tilde{\bc}_1^\pm$ enter the principal Riemann sheet in $p_1'^2$ above the non-anomalous Baikov singularities $\bc_1^\pm$.
This is in contrast to the behaviour in the other anomalous region for this integral~\cite{Lucha:2006vc}.
We point out that the Cauchy principal value operator $\mathcal{P}$ is required here even in the Euclidean region of $p_{1,2}^2<0$ to handle the integrand singularities.
Also, note that $I_{3a}$ contributes only to the real part of the total expression.
By applying the Sokhotski–Plemelj theorem
$\lim_{\varepsilon \to 0^+} (x \pm i \varepsilon)^{-1} = \mathcal{P}(x^{-1}) \mp i \pi \delta(x) \,$ on the non-anomalous contribution $I_3\,$, we checked numerically the new anomalous contribution at $m_3 = m_2$.

%======================================================================
\section{Conclusions and outlook}

In this work, we present a new dispersive formula that retrieves from its maximal cut any multi-point one-loop scalar Feynman integral in the non-anomalous region with unit powers of standard squared propagators and unit numerator.
It requires performing less integrations of an algebraic integrand than other representations and it does not require performing phase space integrals.
We write explicit realization for all examples relevant for collider phenomenology, including the new triple dispersion relation for pentagon and hexagon integrals.
We also find and analyse a new anomalous region for the triangle integral.

A plethora of future directions stems from this work.
For example, it would be interesting to explore beyond-one-loop orders, to systematize the anomalous contributions beyond a triangle, to apply a similar logic in different theories, and finally to apply this technique at the scattering amplitude level within the Multivariate Integrated Unitarity approach.

%======================================================================
\paragraph*{Acknowledgements}

Part of a MSc project contributed to this work. 
We are grateful to GGI for the hospitality during the Mathematical Structures in Scattering Amplitudes 2024 workshop.
We declare that AI was used to understand the multidimensional Baikov positivity region.
We thank G. Sterman for valuable comments on this paper draft.
We are grateful to S. Caron-Huot, M. Correia, E. Gardi, M. Giroux, F. Herren, D. Kermanschah, A. McLeod, S. Mizera, S. Pozzorini, A. Otaifi, K. Sch{\"o}nwald, V. Sotnikov, P. Stoffer, L. Szymanowski, L. Tancredi, M. Zeng, A. Zhiboedov, S. Zoia for interesting discussions. 
The research of PB was supported by the ETH Zurich grant ID030943, the European Research Council (ERC) under the European Union’s Horizon Europe research and innovation program grant agreement 101163627 (ERC Starting Grant ``AmpBoot'').

%======================================================================
\bibliographystyle{apsrev4-1}
\bibliography{references}

\begin{thebibliography}{66}%
\makeatletter
\providecommand \@ifxundefined [1]{%
 \@ifx{#1\undefined}
}%
\providecommand \@ifnum [1]{%
 \ifnum #1\expandafter \@firstoftwo
 \else \expandafter \@secondoftwo
 \fi
}%
\providecommand \@ifx [1]{%
 \ifx #1\expandafter \@firstoftwo
 \else \expandafter \@secondoftwo
 \fi
}%
\providecommand \natexlab [1]{#1}%
\providecommand \enquote  [1]{``#1''}%
\providecommand \bibnamefont  [1]{#1}%
\providecommand \bibfnamefont [1]{#1}%
\providecommand \citenamefont [1]{#1}%
\providecommand \href@noop [0]{\@secondoftwo}%
\providecommand \href [0]{\begingroup \@sanitize@url \@href}%
\providecommand \@href[1]{\@@startlink{#1}\@@href}%
\providecommand \@@href[1]{\endgroup#1\@@endlink}%
\providecommand \@sanitize@url [0]{\catcode `\\12\catcode `\$12\catcode
  `\&12\catcode `\#12\catcode `\^12\catcode `\_12\catcode `\%12\relax}%
\providecommand \@@startlink[1]{}%
\providecommand \@@endlink[0]{}%
\providecommand \url  [0]{\begingroup\@sanitize@url \@url }%
\providecommand \@url [1]{\endgroup\@href {#1}{\urlprefix }}%
\providecommand \urlprefix  [0]{URL }%
\providecommand \Eprint [0]{\href }%
\providecommand \doibase [0]{http://dx.doi.org/}%
\providecommand \selectlanguage [0]{\@gobble}%
\providecommand \bibinfo  [0]{\@secondoftwo}%
\providecommand \bibfield  [0]{\@secondoftwo}%
\providecommand \translation [1]{[#1]}%
\providecommand \BibitemOpen [0]{}%
\providecommand \bibitemStop [0]{}%
\providecommand \bibitemNoStop [0]{.\EOS\space}%
\providecommand \EOS [0]{\spacefactor3000\relax}%
\providecommand \BibitemShut  [1]{\csname bibitem#1\endcsname}%
\let\auto@bib@innerbib\@empty
%</preamble>
\bibitem [{\citenamefont {Heinrich}(2021)}]{Heinrich:2020ybq}%
  \BibitemOpen
  \bibfield  {author} {\bibinfo {author} {\bibfnamefont {G.}~\bibnamefont
  {Heinrich}},\ }\href {\doibase 10.1016/j.physrep.2021.03.006} {\bibfield
  {journal} {\bibinfo  {journal} {Phys. Rept.}\ }\textbf {\bibinfo {volume}
  {922}},\ \bibinfo {pages} {1} (\bibinfo {year} {2021})},\ \Eprint
  {http://arxiv.org/abs/2009.00516} {arXiv:2009.00516 [hep-ph]} \BibitemShut
  {NoStop}%
\bibitem [{\citenamefont {Bourjaily}\ \emph {et~al.}(2022)\citenamefont
  {Bourjaily} \emph {et~al.}}]{Bourjaily:2022bwx}%
  \BibitemOpen
  \bibfield  {author} {\bibinfo {author} {\bibfnamefont {J.~L.}\ \bibnamefont
  {Bourjaily}} \emph {et~al.},\ }in\ \href@noop {} {\emph {\bibinfo {booktitle}
  {{Snowmass 2021}}}}\ (\bibinfo {year} {2022})\ \Eprint
  {http://arxiv.org/abs/2203.07088} {arXiv:2203.07088 [hep-ph]} \BibitemShut
  {NoStop}%
\bibitem [{\citenamefont {Ellis}\ and\ \citenamefont
  {Zanderighi}(2008)}]{Ellis:2007qk}%
  \BibitemOpen
  \bibfield  {author} {\bibinfo {author} {\bibfnamefont {R.~K.}\ \bibnamefont
  {Ellis}}\ and\ \bibinfo {author} {\bibfnamefont {G.}~\bibnamefont
  {Zanderighi}},\ }\href {\doibase 10.1088/1126-6708/2008/02/002} {\bibfield
  {journal} {\bibinfo  {journal} {JHEP}\ }\textbf {\bibinfo {volume} {02}},\
  \bibinfo {pages} {002} (\bibinfo {year} {2008})},\ \Eprint
  {http://arxiv.org/abs/0712.1851} {arXiv:0712.1851 [hep-ph]} \BibitemShut
  {NoStop}%
\bibitem [{\citenamefont {Patel}(2017)}]{Patel:2016fam}%
  \BibitemOpen
  \bibfield  {author} {\bibinfo {author} {\bibfnamefont {H.~H.}\ \bibnamefont
  {Patel}},\ }\href {\doibase 10.1016/j.cpc.2017.04.015} {\bibfield  {journal}
  {\bibinfo  {journal} {Comput. Phys. Commun.}\ }\textbf {\bibinfo {volume}
  {218}},\ \bibinfo {pages} {66} (\bibinfo {year} {2017})},\ \Eprint
  {http://arxiv.org/abs/1612.00009} {arXiv:1612.00009 [hep-ph]} \BibitemShut
  {NoStop}%
\bibitem [{\citenamefont {Panzer}(2015)}]{Panzer:2014caa}%
  \BibitemOpen
  \bibfield  {author} {\bibinfo {author} {\bibfnamefont {E.}~\bibnamefont
  {Panzer}},\ }\href {\doibase 10.1016/j.cpc.2014.10.019} {\bibfield  {journal}
  {\bibinfo  {journal} {Comput. Phys. Commun.}\ }\textbf {\bibinfo {volume}
  {188}},\ \bibinfo {pages} {148} (\bibinfo {year} {2015})},\ \Eprint
  {http://arxiv.org/abs/1403.3385} {arXiv:1403.3385 [hep-th]} \BibitemShut
  {NoStop}%
\bibitem [{\citenamefont {Giroux}\ \emph {et~al.}(2026)\citenamefont {Giroux},
  \citenamefont {Mizera},\ and\ \citenamefont {Salvatori}}]{Giroux:2026tgd}%
  \BibitemOpen
  \bibfield  {author} {\bibinfo {author} {\bibfnamefont {M.}~\bibnamefont
  {Giroux}}, \bibinfo {author} {\bibfnamefont {S.}~\bibnamefont {Mizera}}, \
  and\ \bibinfo {author} {\bibfnamefont {G.}~\bibnamefont {Salvatori}},\
  }\href@noop {} {\  (\bibinfo {year} {2026})},\ \Eprint
  {http://arxiv.org/abs/2604.20954} {arXiv:2604.20954 [hep-th]} \BibitemShut
  {NoStop}%
\bibitem [{\citenamefont {Liu}\ and\ \citenamefont {Ma}(2023)}]{Liu:2022chg}%
  \BibitemOpen
  \bibfield  {author} {\bibinfo {author} {\bibfnamefont {X.}~\bibnamefont
  {Liu}}\ and\ \bibinfo {author} {\bibfnamefont {Y.-Q.}\ \bibnamefont {Ma}},\
  }\href {\doibase 10.1016/j.cpc.2022.108565} {\bibfield  {journal} {\bibinfo
  {journal} {Comput. Phys. Commun.}\ }\textbf {\bibinfo {volume} {283}},\
  \bibinfo {pages} {108565} (\bibinfo {year} {2023})},\ \Eprint
  {http://arxiv.org/abs/2201.11669} {arXiv:2201.11669 [hep-ph]} \BibitemShut
  {NoStop}%
\bibitem [{\citenamefont {Passarino}\ and\ \citenamefont
  {Veltman}(1979)}]{Passarino:1978jh}%
  \BibitemOpen
  \bibfield  {author} {\bibinfo {author} {\bibfnamefont {G.}~\bibnamefont
  {Passarino}}\ and\ \bibinfo {author} {\bibfnamefont {M.~J.~G.}\ \bibnamefont
  {Veltman}},\ }\href {\doibase 10.1016/0550-3213(79)90234-7} {\bibfield
  {journal} {\bibinfo  {journal} {Nucl. Phys. B}\ }\textbf {\bibinfo {volume}
  {160}},\ \bibinfo {pages} {151} (\bibinfo {year} {1979})}\BibitemShut
  {NoStop}%
\bibitem [{\citenamefont {Tkachov}(1981)}]{Tkachov:1981wb}%
  \BibitemOpen
  \bibfield  {author} {\bibinfo {author} {\bibfnamefont {F.~V.}\ \bibnamefont
  {Tkachov}},\ }\href {\doibase 10.1016/0370-2693(81)90288-4} {\bibfield
  {journal} {\bibinfo  {journal} {Phys. Lett. B}\ }\textbf {\bibinfo {volume}
  {100}},\ \bibinfo {pages} {65} (\bibinfo {year} {1981})}\BibitemShut
  {NoStop}%
\bibitem [{\citenamefont {Chetyrkin}\ and\ \citenamefont
  {Tkachov}(1981)}]{Chetyrkin:1981qh}%
  \BibitemOpen
  \bibfield  {author} {\bibinfo {author} {\bibfnamefont {K.~G.}\ \bibnamefont
  {Chetyrkin}}\ and\ \bibinfo {author} {\bibfnamefont {F.~V.}\ \bibnamefont
  {Tkachov}},\ }\href {\doibase 10.1016/0550-3213(81)90199-1} {\bibfield
  {journal} {\bibinfo  {journal} {Nucl. Phys. B}\ }\textbf {\bibinfo {volume}
  {192}},\ \bibinfo {pages} {159} (\bibinfo {year} {1981})}\BibitemShut
  {NoStop}%
\bibitem [{\citenamefont {Bern}\ \emph {et~al.}(1993)\citenamefont {Bern},
  \citenamefont {Dixon},\ and\ \citenamefont {Kosower}}]{Bern:1992em}%
  \BibitemOpen
  \bibfield  {author} {\bibinfo {author} {\bibfnamefont {Z.}~\bibnamefont
  {Bern}}, \bibinfo {author} {\bibfnamefont {L.~J.}\ \bibnamefont {Dixon}}, \
  and\ \bibinfo {author} {\bibfnamefont {D.~A.}\ \bibnamefont {Kosower}},\
  }\href {\doibase 10.1016/0370-2693(93)90400-C} {\bibfield  {journal}
  {\bibinfo  {journal} {Phys. Lett. B}\ }\textbf {\bibinfo {volume} {302}},\
  \bibinfo {pages} {299} (\bibinfo {year} {1993})},\ \bibinfo {note} {[Erratum:
  Phys.Lett.B 318, 649 (1993)]},\ \Eprint {http://arxiv.org/abs/hep-ph/9212308}
  {arXiv:hep-ph/9212308} \BibitemShut {NoStop}%
\bibitem [{\citenamefont {Bern}\ \emph {et~al.}(1994)\citenamefont {Bern},
  \citenamefont {Dixon},\ and\ \citenamefont {Kosower}}]{Bern:1993kr}%
  \BibitemOpen
  \bibfield  {author} {\bibinfo {author} {\bibfnamefont {Z.}~\bibnamefont
  {Bern}}, \bibinfo {author} {\bibfnamefont {L.~J.}\ \bibnamefont {Dixon}}, \
  and\ \bibinfo {author} {\bibfnamefont {D.~A.}\ \bibnamefont {Kosower}},\
  }\href {\doibase 10.1016/0550-3213(94)90398-0} {\bibfield  {journal}
  {\bibinfo  {journal} {Nucl. Phys. B}\ }\textbf {\bibinfo {volume} {412}},\
  \bibinfo {pages} {751} (\bibinfo {year} {1994})},\ \Eprint
  {http://arxiv.org/abs/hep-ph/9306240} {arXiv:hep-ph/9306240} \BibitemShut
  {NoStop}%
\bibitem [{\citenamefont {Bargiela}\ and\ \citenamefont
  {Yang}(2024)}]{Bargiela:2024rul}%
  \BibitemOpen
  \bibfield  {author} {\bibinfo {author} {\bibfnamefont {P.}~\bibnamefont
  {Bargiela}}\ and\ \bibinfo {author} {\bibfnamefont {T.-Z.}\ \bibnamefont
  {Yang}},\ }\href {\doibase 10.1103/PhysRevD.110.096019} {\bibfield  {journal}
  {\bibinfo  {journal} {Phys. Rev. D}\ }\textbf {\bibinfo {volume} {110}},\
  \bibinfo {pages} {096019} (\bibinfo {year} {2024})},\ \Eprint
  {http://arxiv.org/abs/2408.06325} {arXiv:2408.06325 [hep-ph]} \BibitemShut
  {NoStop}%
\bibitem [{\citenamefont {Mandelstam}(1958)}]{Mandelstam:1958xc}%
  \BibitemOpen
  \bibfield  {author} {\bibinfo {author} {\bibfnamefont {S.}~\bibnamefont
  {Mandelstam}},\ }\href {\doibase 10.1103/PhysRev.112.1344} {\bibfield
  {journal} {\bibinfo  {journal} {Phys. Rev.}\ }\textbf {\bibinfo {volume}
  {112}},\ \bibinfo {pages} {1344} (\bibinfo {year} {1958})}\BibitemShut
  {NoStop}%
\bibitem [{\citenamefont
  {Mandelstam}(1959{\natexlab{a}})}]{mandelstam1959construction}%
  \BibitemOpen
  \bibfield  {author} {\bibinfo {author} {\bibfnamefont {S.}~\bibnamefont
  {Mandelstam}},\ }\href@noop {} {\bibfield  {journal} {\bibinfo  {journal}
  {Physical Review}\ }\textbf {\bibinfo {volume} {115}},\ \bibinfo {pages}
  {1752} (\bibinfo {year} {1959}{\natexlab{a}})}\BibitemShut {NoStop}%
\bibitem [{\citenamefont {Mandelstam}(1959{\natexlab{b}})}]{Mandelstam:1959bc}%
  \BibitemOpen
  \bibfield  {author} {\bibinfo {author} {\bibfnamefont {S.}~\bibnamefont
  {Mandelstam}},\ }\href {\doibase 10.1103/PhysRev.115.1741} {\bibfield
  {journal} {\bibinfo  {journal} {Phys. Rev.}\ }\textbf {\bibinfo {volume}
  {115}},\ \bibinfo {pages} {1741} (\bibinfo {year}
  {1959}{\natexlab{b}})}\BibitemShut {NoStop}%
\bibitem [{\citenamefont {Pire}\ \emph {et~al.}(2010)\citenamefont {Pire},
  \citenamefont {Semenov-Tian-Shansky},\ and\ \citenamefont
  {Szymanowski}}]{Pire:2010if}%
  \BibitemOpen
  \bibfield  {author} {\bibinfo {author} {\bibfnamefont {B.}~\bibnamefont
  {Pire}}, \bibinfo {author} {\bibfnamefont {K.}~\bibnamefont
  {Semenov-Tian-Shansky}}, \ and\ \bibinfo {author} {\bibfnamefont
  {L.}~\bibnamefont {Szymanowski}},\ }\href {\doibase
  10.1103/PhysRevD.82.094030} {\bibfield  {journal} {\bibinfo  {journal} {Phys.
  Rev. D}\ }\textbf {\bibinfo {volume} {82}},\ \bibinfo {pages} {094030}
  (\bibinfo {year} {2010})},\ \Eprint {http://arxiv.org/abs/1008.0721}
  {arXiv:1008.0721 [hep-ph]} \BibitemShut {NoStop}%
\bibitem [{\citenamefont {Colangelo}\ \emph {et~al.}(2015)\citenamefont
  {Colangelo}, \citenamefont {Hoferichter}, \citenamefont {Procura},\ and\
  \citenamefont {Stoffer}}]{Colangelo:2015ama}%
  \BibitemOpen
  \bibfield  {author} {\bibinfo {author} {\bibfnamefont {G.}~\bibnamefont
  {Colangelo}}, \bibinfo {author} {\bibfnamefont {M.}~\bibnamefont
  {Hoferichter}}, \bibinfo {author} {\bibfnamefont {M.}~\bibnamefont
  {Procura}}, \ and\ \bibinfo {author} {\bibfnamefont {P.}~\bibnamefont
  {Stoffer}},\ }\href {\doibase 10.1007/JHEP09(2015)074} {\bibfield  {journal}
  {\bibinfo  {journal} {JHEP}\ }\textbf {\bibinfo {volume} {09}},\ \bibinfo
  {pages} {074} (\bibinfo {year} {2015})},\ \Eprint
  {http://arxiv.org/abs/1506.01386} {arXiv:1506.01386 [hep-ph]} \BibitemShut
  {NoStop}%
\bibitem [{\citenamefont {Remiddi}\ and\ \citenamefont
  {Tancredi}(2016)}]{remiddi2016differential}%
  \BibitemOpen
  \bibfield  {author} {\bibinfo {author} {\bibfnamefont {E.}~\bibnamefont
  {Remiddi}}\ and\ \bibinfo {author} {\bibfnamefont {L.}~\bibnamefont
  {Tancredi}},\ }\href@noop {} {\bibfield  {journal} {\bibinfo  {journal}
  {Nuclear Physics B}\ }\textbf {\bibinfo {volume} {907}},\ \bibinfo {pages}
  {400} (\bibinfo {year} {2016})}\BibitemShut {NoStop}%
\bibitem [{\citenamefont {Abreu}\ \emph {et~al.}(2017)\citenamefont {Abreu},
  \citenamefont {Britto}, \citenamefont {Duhr},\ and\ \citenamefont
  {Gardi}}]{abreu2017cuts}%
  \BibitemOpen
  \bibfield  {author} {\bibinfo {author} {\bibfnamefont {S.}~\bibnamefont
  {Abreu}}, \bibinfo {author} {\bibfnamefont {R.}~\bibnamefont {Britto}},
  \bibinfo {author} {\bibfnamefont {C.}~\bibnamefont {Duhr}}, \ and\ \bibinfo
  {author} {\bibfnamefont {E.}~\bibnamefont {Gardi}},\ }\href@noop {}
  {\bibfield  {journal} {\bibinfo  {journal} {Journal of High Energy Physics}\
  }\textbf {\bibinfo {volume} {2017}},\ \bibinfo {pages} {1} (\bibinfo {year}
  {2017})}\BibitemShut {NoStop}%
\bibitem [{\citenamefont {Homrich}\ \emph {et~al.}(2019)\citenamefont
  {Homrich}, \citenamefont {Penedones}, \citenamefont {Toledo}, \citenamefont
  {van Rees},\ and\ \citenamefont {Vieira}}]{Homrich:2019cbt}%
  \BibitemOpen
  \bibfield  {author} {\bibinfo {author} {\bibfnamefont {A.}~\bibnamefont
  {Homrich}}, \bibinfo {author} {\bibfnamefont {J.}~\bibnamefont {Penedones}},
  \bibinfo {author} {\bibfnamefont {J.}~\bibnamefont {Toledo}}, \bibinfo
  {author} {\bibfnamefont {B.~C.}\ \bibnamefont {van Rees}}, \ and\ \bibinfo
  {author} {\bibfnamefont {P.}~\bibnamefont {Vieira}},\ }\href {\doibase
  10.1007/JHEP11(2019)076} {\bibfield  {journal} {\bibinfo  {journal} {JHEP}\
  }\textbf {\bibinfo {volume} {11}},\ \bibinfo {pages} {076} (\bibinfo {year}
  {2019})},\ \Eprint {http://arxiv.org/abs/1905.06905} {arXiv:1905.06905
  [hep-th]} \BibitemShut {NoStop}%
\bibitem [{\citenamefont {Correia}\ \emph {et~al.}(2021)\citenamefont
  {Correia}, \citenamefont {Sever},\ and\ \citenamefont
  {Zhiboedov}}]{Correia:2020xtr}%
  \BibitemOpen
  \bibfield  {author} {\bibinfo {author} {\bibfnamefont {M.}~\bibnamefont
  {Correia}}, \bibinfo {author} {\bibfnamefont {A.}~\bibnamefont {Sever}}, \
  and\ \bibinfo {author} {\bibfnamefont {A.}~\bibnamefont {Zhiboedov}},\ }\href
  {\doibase 10.1007/JHEP03(2021)013} {\bibfield  {journal} {\bibinfo  {journal}
  {JHEP}\ }\textbf {\bibinfo {volume} {03}},\ \bibinfo {pages} {013} (\bibinfo
  {year} {2021})},\ \Eprint {http://arxiv.org/abs/2006.08221} {arXiv:2006.08221
  [hep-th]} \BibitemShut {NoStop}%
\bibitem [{\citenamefont {Correia}(2024)}]{Correia:2022dcu}%
  \BibitemOpen
  \bibfield  {author} {\bibinfo {author} {\bibfnamefont {M.}~\bibnamefont
  {Correia}},\ }\href {\doibase 10.1103/PhysRevD.110.025012} {\bibfield
  {journal} {\bibinfo  {journal} {Phys. Rev. D}\ }\textbf {\bibinfo {volume}
  {110}},\ \bibinfo {pages} {025012} (\bibinfo {year} {2024})},\ \Eprint
  {http://arxiv.org/abs/2212.06157} {arXiv:2212.06157 [hep-th]} \BibitemShut
  {NoStop}%
\bibitem [{\citenamefont {Vergu}(2024)}]{Vergu:2023rqz}%
  \BibitemOpen
  \bibfield  {author} {\bibinfo {author} {\bibfnamefont {C.}~\bibnamefont
  {Vergu}},\ }\href {\doibase 10.1007/JHEP05(2024)302} {\bibfield  {journal}
  {\bibinfo  {journal} {JHEP}\ }\textbf {\bibinfo {volume} {05}},\ \bibinfo
  {pages} {302} (\bibinfo {year} {2024})},\ \Eprint
  {http://arxiv.org/abs/2311.16069} {arXiv:2311.16069 [hep-th]} \BibitemShut
  {NoStop}%
\bibitem [{\citenamefont {Tourkine}\ and\ \citenamefont
  {Zhiboedov}(2023)}]{Tourkine:2023xtu}%
  \BibitemOpen
  \bibfield  {author} {\bibinfo {author} {\bibfnamefont {P.}~\bibnamefont
  {Tourkine}}\ and\ \bibinfo {author} {\bibfnamefont {A.}~\bibnamefont
  {Zhiboedov}},\ }\href {\doibase 10.1007/JHEP11(2023)005} {\bibfield
  {journal} {\bibinfo  {journal} {JHEP}\ }\textbf {\bibinfo {volume} {11}},\
  \bibinfo {pages} {005} (\bibinfo {year} {2023})},\ \Eprint
  {http://arxiv.org/abs/2303.08839} {arXiv:2303.08839 [hep-th]} \BibitemShut
  {NoStop}%
\bibitem [{\citenamefont {Dawid}\ \emph {et~al.}(2025)\citenamefont {Dawid},
  \citenamefont {Jackura},\ and\ \citenamefont {Szczepaniak}}]{Dawid:2024oey}%
  \BibitemOpen
  \bibfield  {author} {\bibinfo {author} {\bibfnamefont {S.~M.}\ \bibnamefont
  {Dawid}}, \bibinfo {author} {\bibfnamefont {A.~W.}\ \bibnamefont {Jackura}},
  \ and\ \bibinfo {author} {\bibfnamefont {A.~P.}\ \bibnamefont
  {Szczepaniak}},\ }\href {\doibase 10.1016/j.physletb.2025.139442} {\bibfield
  {journal} {\bibinfo  {journal} {Phys. Lett. B}\ }\textbf {\bibinfo {volume}
  {864}},\ \bibinfo {pages} {139442} (\bibinfo {year} {2025})},\ \Eprint
  {http://arxiv.org/abs/2411.15730} {arXiv:2411.15730 [hep-lat]} \BibitemShut
  {NoStop}%
\bibitem [{\citenamefont {Hoferichter}\ \emph {et~al.}(2025)\citenamefont
  {Hoferichter}, \citenamefont {Stoffer},\ and\ \citenamefont
  {Zillinger}}]{Hoferichter:2024vbu}%
  \BibitemOpen
  \bibfield  {author} {\bibinfo {author} {\bibfnamefont {M.}~\bibnamefont
  {Hoferichter}}, \bibinfo {author} {\bibfnamefont {P.}~\bibnamefont
  {Stoffer}}, \ and\ \bibinfo {author} {\bibfnamefont {M.}~\bibnamefont
  {Zillinger}},\ }\href {\doibase 10.1103/PhysRevLett.134.061902} {\bibfield
  {journal} {\bibinfo  {journal} {Phys. Rev. Lett.}\ }\textbf {\bibinfo
  {volume} {134}},\ \bibinfo {pages} {061902} (\bibinfo {year} {2025})},\
  \Eprint {http://arxiv.org/abs/2412.00190} {arXiv:2412.00190 [hep-ph]}
  \BibitemShut {NoStop}%
\bibitem [{\citenamefont {Herren}\ \emph {et~al.}(2026)\citenamefont {Herren},
  \citenamefont {Nouri},\ and\ \citenamefont {Reboud}}]{Herren:2026xfq}%
  \BibitemOpen
  \bibfield  {author} {\bibinfo {author} {\bibfnamefont {F.}~\bibnamefont
  {Herren}}, \bibinfo {author} {\bibfnamefont {F.}~\bibnamefont {Nouri}}, \
  and\ \bibinfo {author} {\bibfnamefont {M.}~\bibnamefont {Reboud}},\
  }\href@noop {} {\  (\bibinfo {year} {2026})},\ \Eprint
  {http://arxiv.org/abs/2611.16974} {arXiv:2611.16974 [hep-ph]} \BibitemShut
  {NoStop}%
\bibitem [{\citenamefont {Eden}\ \emph {et~al.}(1966)\citenamefont {Eden},
  \citenamefont {Landshoff}, \citenamefont {Olive},\ and\ \citenamefont
  {Polkinghorne}}]{Eden:1966dnq}%
  \BibitemOpen
  \bibfield  {author} {\bibinfo {author} {\bibfnamefont {R.~J.}\ \bibnamefont
  {Eden}}, \bibinfo {author} {\bibfnamefont {P.~V.}\ \bibnamefont {Landshoff}},
  \bibinfo {author} {\bibfnamefont {D.~I.}\ \bibnamefont {Olive}}, \ and\
  \bibinfo {author} {\bibfnamefont {J.~C.}\ \bibnamefont {Polkinghorne}},\
  }\href@noop {} {\emph {\bibinfo {title} {{The analytic S-matrix}}}}\
  (\bibinfo  {publisher} {Cambridge Univ. Press},\ \bibinfo {address}
  {Cambridge},\ \bibinfo {year} {1966})\BibitemShut {NoStop}%
\bibitem [{\citenamefont {Hannesdottir}\ \emph {et~al.}(2025)\citenamefont
  {Hannesdottir}, \citenamefont {McLeod}, \citenamefont {Schwartz},\ and\
  \citenamefont {Vergu}}]{Hannesdottir:2024hke}%
  \BibitemOpen
  \bibfield  {author} {\bibinfo {author} {\bibfnamefont {H.~S.}\ \bibnamefont
  {Hannesdottir}}, \bibinfo {author} {\bibfnamefont {A.~J.}\ \bibnamefont
  {McLeod}}, \bibinfo {author} {\bibfnamefont {M.~D.}\ \bibnamefont
  {Schwartz}}, \ and\ \bibinfo {author} {\bibfnamefont {C.}~\bibnamefont
  {Vergu}},\ }\href {\doibase 10.1103/PhysRevD.111.085003} {\bibfield
  {journal} {\bibinfo  {journal} {Phys. Rev. D}\ }\textbf {\bibinfo {volume}
  {111}},\ \bibinfo {pages} {085003} (\bibinfo {year} {2025})},\ \Eprint
  {http://arxiv.org/abs/2410.02424} {arXiv:2410.02424 [hep-ph]} \BibitemShut
  {NoStop}%
\bibitem [{\citenamefont {McLeod}\ \emph {et~al.}(2026)\citenamefont {McLeod},
  \citenamefont {Bargie{\l}a}, \citenamefont {Crisanti}, \citenamefont
  {Larkin}, \citenamefont {Lippstreu}, \citenamefont {Polackova},\ and\
  \citenamefont {Walsh}}]{McLeod:2026kaf}%
  \BibitemOpen
  \bibfield  {author} {\bibinfo {author} {\bibfnamefont {A.~J.}\ \bibnamefont
  {McLeod}}, \bibinfo {author} {\bibfnamefont {P.}~\bibnamefont {Bargie{\l}a}},
  \bibinfo {author} {\bibfnamefont {G.}~\bibnamefont {Crisanti}}, \bibinfo
  {author} {\bibfnamefont {C.}~\bibnamefont {Larkin}}, \bibinfo {author}
  {\bibfnamefont {L.}~\bibnamefont {Lippstreu}}, \bibinfo {author}
  {\bibfnamefont {M.}~\bibnamefont {Polackova}}, \ and\ \bibinfo {author}
  {\bibfnamefont {L.}~\bibnamefont {Walsh}},\ }\href {\doibase
  10.22323/1.521.0068} {\bibfield  {journal} {\bibinfo  {journal} {PoS}\
  }\textbf {\bibinfo {volume} {LL2026}},\ \bibinfo {pages} {068} (\bibinfo
  {year} {2026})},\ \Eprint {http://arxiv.org/abs/2607.16034} {arXiv:2607.16034
  [hep-th]} \BibitemShut {NoStop}%
\bibitem [{\citenamefont {Landau}(1960)}]{Landau:1959fi}%
  \BibitemOpen
  \bibfield  {author} {\bibinfo {author} {\bibfnamefont {L.~D.}\ \bibnamefont
  {Landau}},\ }\href {\doibase 10.1016/B978-0-08-010586-4.50103-6} {\bibfield
  {journal} {\bibinfo  {journal} {Zh. Eksp. Teor. Fiz.}\ }\textbf {\bibinfo
  {volume} {37}},\ \bibinfo {pages} {62} (\bibinfo {year} {1960})}\BibitemShut
  {NoStop}%
\bibitem [{\citenamefont {Correia}\ \emph {et~al.}(2025)\citenamefont
  {Correia}, \citenamefont {Giroux},\ and\ \citenamefont
  {Mizera}}]{Correia:2025yao}%
  \BibitemOpen
  \bibfield  {author} {\bibinfo {author} {\bibfnamefont {M.}~\bibnamefont
  {Correia}}, \bibinfo {author} {\bibfnamefont {M.}~\bibnamefont {Giroux}}, \
  and\ \bibinfo {author} {\bibfnamefont {S.}~\bibnamefont {Mizera}},\
  }\href@noop {} {\  (\bibinfo {year} {2025})},\ \Eprint
  {http://arxiv.org/abs/2503.16601} {arXiv:2503.16601 [hep-th]} \BibitemShut
  {NoStop}%
\bibitem [{\citenamefont {Correia}\ \emph {et~al.}(2022)\citenamefont
  {Correia}, \citenamefont {Sever},\ and\ \citenamefont
  {Zhiboedov}}]{Correia:2021etg}%
  \BibitemOpen
  \bibfield  {author} {\bibinfo {author} {\bibfnamefont {M.}~\bibnamefont
  {Correia}}, \bibinfo {author} {\bibfnamefont {A.}~\bibnamefont {Sever}}, \
  and\ \bibinfo {author} {\bibfnamefont {A.}~\bibnamefont {Zhiboedov}},\ }\href
  {\doibase 10.21468/SciPostPhys.13.3.062} {\bibfield  {journal} {\bibinfo
  {journal} {SciPost Phys.}\ }\textbf {\bibinfo {volume} {13}},\ \bibinfo
  {pages} {062} (\bibinfo {year} {2022})},\ \Eprint
  {http://arxiv.org/abs/2111.12100} {arXiv:2111.12100 [hep-th]} \BibitemShut
  {NoStop}%
\bibitem [{\citenamefont {Hannesdottir}\ and\ \citenamefont
  {Mizera}(2023)}]{Hannesdottir:2022bmo}%
  \BibitemOpen
  \bibfield  {author} {\bibinfo {author} {\bibfnamefont {H.~S.}\ \bibnamefont
  {Hannesdottir}}\ and\ \bibinfo {author} {\bibfnamefont {S.}~\bibnamefont
  {Mizera}},\ }\href {\doibase 10.1007/978-3-031-18258-7} {\emph {\bibinfo
  {title} {{What is the i{\ensuremath{\varepsilon}} for the S-matrix?}}}},\
  SpringerBriefs in Physics\ (\bibinfo  {publisher} {Springer},\ \bibinfo
  {year} {2023})\ \Eprint {http://arxiv.org/abs/2204.02988} {arXiv:2204.02988
  [hep-th]} \BibitemShut {NoStop}%
\bibitem [{\citenamefont {Caron-Huot}\ \emph {et~al.}(2024)\citenamefont
  {Caron-Huot}, \citenamefont {Giroux}, \citenamefont {Hannesdottir},\ and\
  \citenamefont {Mizera}}]{Caron-Huot:2023ikn}%
  \BibitemOpen
  \bibfield  {author} {\bibinfo {author} {\bibfnamefont {S.}~\bibnamefont
  {Caron-Huot}}, \bibinfo {author} {\bibfnamefont {M.}~\bibnamefont {Giroux}},
  \bibinfo {author} {\bibfnamefont {H.~S.}\ \bibnamefont {Hannesdottir}}, \
  and\ \bibinfo {author} {\bibfnamefont {S.}~\bibnamefont {Mizera}},\ }\href
  {\doibase 10.1007/JHEP04(2024)060} {\bibfield  {journal} {\bibinfo  {journal}
  {JHEP}\ }\textbf {\bibinfo {volume} {04}},\ \bibinfo {pages} {060} (\bibinfo
  {year} {2024})},\ \Eprint {http://arxiv.org/abs/2310.12199} {arXiv:2310.12199
  [hep-th]} \BibitemShut {NoStop}%
\bibitem [{\citenamefont {Bargiela}(2025)}]{Bargiela:2024hgt}%
  \BibitemOpen
  \bibfield  {author} {\bibinfo {author} {\bibfnamefont {P.}~\bibnamefont
  {Bargiela}},\ }\href {\doibase 10.1007/JHEP05(2025)004} {\bibfield  {journal}
  {\bibinfo  {journal} {JHEP}\ }\textbf {\bibinfo {volume} {05}},\ \bibinfo
  {pages} {004} (\bibinfo {year} {2025})},\ \Eprint
  {http://arxiv.org/abs/2403.18047} {arXiv:2403.18047 [hep-th]} \BibitemShut
  {NoStop}%
\bibitem [{\citenamefont {Britto}\ \emph {et~al.}(2005)\citenamefont {Britto},
  \citenamefont {Cachazo},\ and\ \citenamefont {Feng}}]{Britto:2004nc}%
  \BibitemOpen
  \bibfield  {author} {\bibinfo {author} {\bibfnamefont {R.}~\bibnamefont
  {Britto}}, \bibinfo {author} {\bibfnamefont {F.}~\bibnamefont {Cachazo}}, \
  and\ \bibinfo {author} {\bibfnamefont {B.}~\bibnamefont {Feng}},\ }\href
  {\doibase 10.1016/j.nuclphysb.2005.07.014} {\bibfield  {journal} {\bibinfo
  {journal} {Nucl. Phys. B}\ }\textbf {\bibinfo {volume} {725}},\ \bibinfo
  {pages} {275} (\bibinfo {year} {2005})},\ \Eprint
  {http://arxiv.org/abs/hep-th/0412103} {arXiv:hep-th/0412103} \BibitemShut
  {NoStop}%
\bibitem [{\citenamefont {Anastasiou}\ \emph {et~al.}(2007)\citenamefont
  {Anastasiou}, \citenamefont {Britto}, \citenamefont {Feng}, \citenamefont
  {Kunszt},\ and\ \citenamefont {Mastrolia}}]{Anastasiou:2006jv}%
  \BibitemOpen
  \bibfield  {author} {\bibinfo {author} {\bibfnamefont {C.}~\bibnamefont
  {Anastasiou}}, \bibinfo {author} {\bibfnamefont {R.}~\bibnamefont {Britto}},
  \bibinfo {author} {\bibfnamefont {B.}~\bibnamefont {Feng}}, \bibinfo {author}
  {\bibfnamefont {Z.}~\bibnamefont {Kunszt}}, \ and\ \bibinfo {author}
  {\bibfnamefont {P.}~\bibnamefont {Mastrolia}},\ }\href {\doibase
  10.1016/j.physletb.2006.12.022} {\bibfield  {journal} {\bibinfo  {journal}
  {Phys. Lett. B}\ }\textbf {\bibinfo {volume} {645}},\ \bibinfo {pages} {213}
  (\bibinfo {year} {2007})},\ \Eprint {http://arxiv.org/abs/hep-ph/0609191}
  {arXiv:hep-ph/0609191} \BibitemShut {NoStop}%
\bibitem [{\citenamefont {Forde}(2007)}]{Forde:2007mi}%
  \BibitemOpen
  \bibfield  {author} {\bibinfo {author} {\bibfnamefont {D.}~\bibnamefont
  {Forde}},\ }\href {\doibase 10.1103/PhysRevD.75.125019} {\bibfield  {journal}
  {\bibinfo  {journal} {Phys. Rev. D}\ }\textbf {\bibinfo {volume} {75}},\
  \bibinfo {pages} {125019} (\bibinfo {year} {2007})},\ \Eprint
  {http://arxiv.org/abs/0704.1835} {arXiv:0704.1835 [hep-ph]} \BibitemShut
  {NoStop}%
\bibitem [{\citenamefont {Cutkosky}(1960)}]{Cutkosky:1960sp}%
  \BibitemOpen
  \bibfield  {author} {\bibinfo {author} {\bibfnamefont {R.~E.}\ \bibnamefont
  {Cutkosky}},\ }\href {\doibase 10.1063/1.1703676} {\bibfield  {journal}
  {\bibinfo  {journal} {J. Math. Phys.}\ }\textbf {\bibinfo {volume} {1}},\
  \bibinfo {pages} {429} (\bibinfo {year} {1960})}\BibitemShut {NoStop}%
\bibitem [{\citenamefont {Bosma}\ \emph {et~al.}(2017)\citenamefont {Bosma},
  \citenamefont {Sogaard},\ and\ \citenamefont {Zhang}}]{Bosma:2017ens}%
  \BibitemOpen
  \bibfield  {author} {\bibinfo {author} {\bibfnamefont {J.}~\bibnamefont
  {Bosma}}, \bibinfo {author} {\bibfnamefont {M.}~\bibnamefont {Sogaard}}, \
  and\ \bibinfo {author} {\bibfnamefont {Y.}~\bibnamefont {Zhang}},\ }\href
  {\doibase 10.1007/JHEP08(2017)051} {\bibfield  {journal} {\bibinfo  {journal}
  {JHEP}\ }\textbf {\bibinfo {volume} {08}},\ \bibinfo {pages} {051} (\bibinfo
  {year} {2017})},\ \Eprint {http://arxiv.org/abs/1704.04255} {arXiv:1704.04255
  [hep-th]} \BibitemShut {NoStop}%
\bibitem [{\citenamefont {Harley}\ \emph {et~al.}(2017)\citenamefont {Harley},
  \citenamefont {Moriello},\ and\ \citenamefont {Schabinger}}]{Harley:2017qut}%
  \BibitemOpen
  \bibfield  {author} {\bibinfo {author} {\bibfnamefont {M.}~\bibnamefont
  {Harley}}, \bibinfo {author} {\bibfnamefont {F.}~\bibnamefont {Moriello}}, \
  and\ \bibinfo {author} {\bibfnamefont {R.~M.}\ \bibnamefont {Schabinger}},\
  }\href {\doibase 10.1007/JHEP06(2017)049} {\bibfield  {journal} {\bibinfo
  {journal} {JHEP}\ }\textbf {\bibinfo {volume} {06}},\ \bibinfo {pages} {049}
  (\bibinfo {year} {2017})},\ \Eprint {http://arxiv.org/abs/1705.03478}
  {arXiv:1705.03478 [hep-ph]} \BibitemShut {NoStop}%
\bibitem [{\citenamefont {Baikov}(1997)}]{Baikov:1996iu}%
  \BibitemOpen
  \bibfield  {author} {\bibinfo {author} {\bibfnamefont {P.~A.}\ \bibnamefont
  {Baikov}},\ }\href {\doibase 10.1016/S0168-9002(97)00126-5} {\bibfield
  {journal} {\bibinfo  {journal} {Nucl. Instrum. Meth. A}\ }\textbf {\bibinfo
  {volume} {389}},\ \bibinfo {pages} {347} (\bibinfo {year} {1997})},\ \Eprint
  {http://arxiv.org/abs/hep-ph/9611449} {arXiv:hep-ph/9611449} \BibitemShut
  {NoStop}%
\bibitem [{\citenamefont {Davydychev}(1992)}]{davydychev1992general}%
  \BibitemOpen
  \bibfield  {author} {\bibinfo {author} {\bibfnamefont {A.~I.}\ \bibnamefont
  {Davydychev}},\ }\href@noop {} {\bibfield  {journal} {\bibinfo  {journal}
  {Journal of mathematical physics}\ }\textbf {\bibinfo {volume} {33}},\
  \bibinfo {pages} {358} (\bibinfo {year} {1992})}\BibitemShut {NoStop}%
\bibitem [{\citenamefont {Shpot}(2007)}]{shpot2007massive}%
  \BibitemOpen
  \bibfield  {author} {\bibinfo {author} {\bibfnamefont {M.}~\bibnamefont
  {Shpot}},\ }\href@noop {} {\bibfield  {journal} {\bibinfo  {journal} {Journal
  of mathematical physics}\ }\textbf {\bibinfo {volume} {48}} (\bibinfo {year}
  {2007})}\BibitemShut {NoStop}%
\bibitem [{\citenamefont {Anisovich}\ \emph {et~al.}(1993)\citenamefont
  {Anisovich}, \citenamefont {Melikhov}, \citenamefont {Metsch},\ and\
  \citenamefont {Petry}}]{anisovich1993bethe}%
  \BibitemOpen
  \bibfield  {author} {\bibinfo {author} {\bibfnamefont {V.}~\bibnamefont
  {Anisovich}}, \bibinfo {author} {\bibfnamefont {D.}~\bibnamefont {Melikhov}},
  \bibinfo {author} {\bibfnamefont {B.~C.}\ \bibnamefont {Metsch}}, \ and\
  \bibinfo {author} {\bibfnamefont {H.~R.}\ \bibnamefont {Petry}},\ }\href@noop
  {} {\bibfield  {journal} {\bibinfo  {journal} {Nuclear Physics A}\ }\textbf
  {\bibinfo {volume} {563}},\ \bibinfo {pages} {549} (\bibinfo {year}
  {1993})}\BibitemShut {NoStop}%
\bibitem [{\citenamefont {Fronsdal}\ and\ \citenamefont
  {Norton}(1964{\natexlab{a}})}]{fronsdal1964integral}%
  \BibitemOpen
  \bibfield  {author} {\bibinfo {author} {\bibfnamefont {C.}~\bibnamefont
  {Fronsdal}}\ and\ \bibinfo {author} {\bibfnamefont {R.}~\bibnamefont
  {Norton}},\ }\href@noop {} {\bibfield  {journal} {\bibinfo  {journal}
  {Journal of Mathematical Physics}\ }\textbf {\bibinfo {volume} {5}},\
  \bibinfo {pages} {100} (\bibinfo {year} {1964}{\natexlab{a}})}\BibitemShut
  {NoStop}%
\bibitem [{\citenamefont {Ball}\ \emph {et~al.}(1991)\citenamefont {Ball},
  \citenamefont {Braun},\ and\ \citenamefont {Dosch}}]{ball1991form}%
  \BibitemOpen
  \bibfield  {author} {\bibinfo {author} {\bibfnamefont {P.}~\bibnamefont
  {Ball}}, \bibinfo {author} {\bibfnamefont {V.~M.}\ \bibnamefont {Braun}}, \
  and\ \bibinfo {author} {\bibfnamefont {H.~G.}\ \bibnamefont {Dosch}},\
  }\href@noop {} {\bibfield  {journal} {\bibinfo  {journal} {Physical Review
  D}\ }\textbf {\bibinfo {volume} {44}},\ \bibinfo {pages} {3567} (\bibinfo
  {year} {1991})}\BibitemShut {NoStop}%
\bibitem [{\citenamefont {Coriano}(1995)}]{Coriano:1994sz}%
  \BibitemOpen
  \bibfield  {author} {\bibinfo {author} {\bibfnamefont {C.}~\bibnamefont
  {Coriano}},\ }\href {\doibase 10.1016/0550-3213(94)00486-X} {\bibfield
  {journal} {\bibinfo  {journal} {Nucl. Phys. B}\ }\textbf {\bibinfo {volume}
  {434}},\ \bibinfo {pages} {565} (\bibinfo {year} {1995})},\ \Eprint
  {http://arxiv.org/abs/hep-ph/9405403} {arXiv:hep-ph/9405403} \BibitemShut
  {NoStop}%
\bibitem [{\citenamefont {Lucha}\ \emph {et~al.}(2007)\citenamefont {Lucha},
  \citenamefont {Melikhov},\ and\ \citenamefont {Simula}}]{Lucha:2006vc}%
  \BibitemOpen
  \bibfield  {author} {\bibinfo {author} {\bibfnamefont {W.}~\bibnamefont
  {Lucha}}, \bibinfo {author} {\bibfnamefont {D.}~\bibnamefont {Melikhov}}, \
  and\ \bibinfo {author} {\bibfnamefont {S.}~\bibnamefont {Simula}},\ }\href
  {\doibase 10.1103/PhysRevD.75.016001} {\bibfield  {journal} {\bibinfo
  {journal} {Phys. Rev. D}\ }\textbf {\bibinfo {volume} {75}},\ \bibinfo
  {pages} {016001} (\bibinfo {year} {2007})},\ \bibinfo {note} {[Erratum:
  Phys.Rev.D 92, 019901 (2015)]},\ \Eprint
  {http://arxiv.org/abs/hep-ph/0610330} {arXiv:hep-ph/0610330} \BibitemShut
  {NoStop}%
\bibitem [{\citenamefont {Suzuki}\ \emph {et~al.}(2008)\citenamefont {Suzuki},
  \citenamefont {Schmidt},\ and\ \citenamefont {Bolzan}}]{Suzuki:2008hm}%
  \BibitemOpen
  \bibfield  {author} {\bibinfo {author} {\bibfnamefont {A.~T.}\ \bibnamefont
  {Suzuki}}, \bibinfo {author} {\bibfnamefont {A.~G.~M.}\ \bibnamefont
  {Schmidt}}, \ and\ \bibinfo {author} {\bibfnamefont {J.~D.}\ \bibnamefont
  {Bolzan}},\ }\href@noop {} {\  (\bibinfo {year} {2008})},\ \Eprint
  {http://arxiv.org/abs/0808.1496} {arXiv:0808.1496 [hep-ph]} \BibitemShut
  {NoStop}%
\bibitem [{\citenamefont {De~Tollis}(1964)}]{DeTollis:1964una}%
  \BibitemOpen
  \bibfield  {author} {\bibinfo {author} {\bibfnamefont {B.}~\bibnamefont
  {De~Tollis}},\ }\href {\doibase 10.1007/BF02735895} {\bibfield  {journal}
  {\bibinfo  {journal} {Nuovo Cim.}\ }\textbf {\bibinfo {volume} {32}},\
  \bibinfo {pages} {757} (\bibinfo {year} {1964})}\BibitemShut {NoStop}%
\bibitem [{\citenamefont {JS}\ \emph {et~al.}(1970)\citenamefont {JS},
  \citenamefont {Segerlund},\ and\ \citenamefont {Woolcock}}]{js1970analytic}%
  \BibitemOpen
  \bibfield  {author} {\bibinfo {author} {\bibfnamefont {F.}~\bibnamefont
  {JS}}, \bibinfo {author} {\bibfnamefont {J.}~\bibnamefont {Segerlund}}, \
  and\ \bibinfo {author} {\bibfnamefont {W.}~\bibnamefont {Woolcock}},\
  }\href@noop {} {\emph {\bibinfo {title} {The Analytic Properties of the Box
  Diagram Amplitude: I}}}\ (\bibinfo  {publisher} {Research School of Physical
  Sciences, Australian National University},\ \bibinfo {year}
  {1970})\BibitemShut {NoStop}%
\bibitem [{\citenamefont {Frederiksen}\ and\ \citenamefont
  {Woolcock}(1973)}]{frederiksen1973analytic}%
  \BibitemOpen
  \bibfield  {author} {\bibinfo {author} {\bibfnamefont {J.}~\bibnamefont
  {Frederiksen}}\ and\ \bibinfo {author} {\bibfnamefont {W.}~\bibnamefont
  {Woolcock}},\ }\href@noop {} {\bibfield  {journal} {\bibinfo  {journal}
  {Annals of Physics}\ }\textbf {\bibinfo {volume} {80}},\ \bibinfo {pages}
  {86} (\bibinfo {year} {1973})}\BibitemShut {NoStop}%
\bibitem [{\citenamefont {Duplancic}\ and\ \citenamefont
  {Nizic}(2002)}]{Duplancic:2002dh}%
  \BibitemOpen
  \bibfield  {author} {\bibinfo {author} {\bibfnamefont {G.}~\bibnamefont
  {Duplancic}}\ and\ \bibinfo {author} {\bibfnamefont {B.}~\bibnamefont
  {Nizic}},\ }\href {\doibase 10.1007/s100520200943} {\bibfield  {journal}
  {\bibinfo  {journal} {Eur. Phys. J. C}\ }\textbf {\bibinfo {volume} {24}},\
  \bibinfo {pages} {385} (\bibinfo {year} {2002})},\ \Eprint
  {http://arxiv.org/abs/hep-ph/0201306} {arXiv:hep-ph/0201306} \BibitemShut
  {NoStop}%
\bibitem [{\citenamefont {Denner}\ and\ \citenamefont
  {Dittmaier}(2011)}]{Denner:2010tr}%
  \BibitemOpen
  \bibfield  {author} {\bibinfo {author} {\bibfnamefont {A.}~\bibnamefont
  {Denner}}\ and\ \bibinfo {author} {\bibfnamefont {S.}~\bibnamefont
  {Dittmaier}},\ }\href {\doibase 10.1016/j.nuclphysb.2010.11.002} {\bibfield
  {journal} {\bibinfo  {journal} {Nucl. Phys. B}\ }\textbf {\bibinfo {volume}
  {844}},\ \bibinfo {pages} {199} (\bibinfo {year} {2011})},\ \Eprint
  {http://arxiv.org/abs/1005.2076} {arXiv:1005.2076 [hep-ph]} \BibitemShut
  {NoStop}%
\bibitem [{\citenamefont {Zhang}\ and\ \citenamefont
  {Guo}(2026)}]{Zhang:2026eez}%
  \BibitemOpen
  \bibfield  {author} {\bibinfo {author} {\bibfnamefont {X.}~\bibnamefont
  {Zhang}}\ and\ \bibinfo {author} {\bibfnamefont {F.-K.}\ \bibnamefont
  {Guo}},\ }\href@noop {} {\  (\bibinfo {year} {2026})},\ \Eprint
  {http://arxiv.org/abs/2609.00985} {arXiv:2609.00985 [hep-ph]} \BibitemShut
  {NoStop}%
\bibitem [{\citenamefont {Karplus}\ \emph {et~al.}(1958)\citenamefont
  {Karplus}, \citenamefont {Sommerfield},\ and\ \citenamefont
  {Wichmann}}]{PhysRev.111.1187}%
  \BibitemOpen
  \bibfield  {author} {\bibinfo {author} {\bibfnamefont {R.}~\bibnamefont
  {Karplus}}, \bibinfo {author} {\bibfnamefont {C.~M.}\ \bibnamefont
  {Sommerfield}}, \ and\ \bibinfo {author} {\bibfnamefont {E.~H.}\ \bibnamefont
  {Wichmann}},\ }\href {\doibase 10.1103/PhysRev.111.1187} {\bibfield
  {journal} {\bibinfo  {journal} {Phys. Rev.}\ }\textbf {\bibinfo {volume}
  {111}},\ \bibinfo {pages} {1187} (\bibinfo {year} {1958})}\BibitemShut
  {NoStop}%
\bibitem [{\citenamefont {Mandelstam}(1960)}]{Mandelstam:1960zz}%
  \BibitemOpen
  \bibfield  {author} {\bibinfo {author} {\bibfnamefont {S.}~\bibnamefont
  {Mandelstam}},\ }\href {\doibase 10.1103/PhysRevLett.4.84} {\bibfield
  {journal} {\bibinfo  {journal} {Phys. Rev. Lett.}\ }\textbf {\bibinfo
  {volume} {4}},\ \bibinfo {pages} {84} (\bibinfo {year} {1960})}\BibitemShut
  {NoStop}%
\bibitem [{\citenamefont {Cutkosky}(1961)}]{RevModPhys.33.448}%
  \BibitemOpen
  \bibfield  {author} {\bibinfo {author} {\bibfnamefont {R.~E.}\ \bibnamefont
  {Cutkosky}},\ }\href {\doibase 10.1103/RevModPhys.33.448} {\bibfield
  {journal} {\bibinfo  {journal} {Rev. Mod. Phys.}\ }\textbf {\bibinfo {volume}
  {33}},\ \bibinfo {pages} {448} (\bibinfo {year} {1961})}\BibitemShut
  {NoStop}%
\bibitem [{\citenamefont {Islam}(1962)}]{10.1063/1.1703852}%
  \BibitemOpen
  \bibfield  {author} {\bibinfo {author} {\bibfnamefont {J.~N.}\ \bibnamefont
  {Islam}},\ }\href {\doibase 10.1063/1.1703852} {\bibfield  {journal}
  {\bibinfo  {journal} {Journal of Mathematical Physics}\ }\textbf {\bibinfo
  {volume} {3}},\ \bibinfo {pages} {1098} (\bibinfo {year} {1962})},\ \Eprint
  {http://arxiv.org/abs/https://pubs.aip.org/aip/jmp/article-pdf/3/6/1098/19064301/1098\_1\_online.pdf}
  {https://pubs.aip.org/aip/jmp/article-pdf/3/6/1098/19064301/1098\_1\_online.pdf}
  \BibitemShut {NoStop}%
\bibitem [{\citenamefont {Fronsdal}\ and\ \citenamefont
  {Norton}(1964{\natexlab{b}})}]{10.1063/1.1704053}%
  \BibitemOpen
  \bibfield  {author} {\bibinfo {author} {\bibfnamefont {C.}~\bibnamefont
  {Fronsdal}}\ and\ \bibinfo {author} {\bibfnamefont {R.~E.}\ \bibnamefont
  {Norton}},\ }\href {\doibase 10.1063/1.1704053} {\bibfield  {journal}
  {\bibinfo  {journal} {Journal of Mathematical Physics}\ }\textbf {\bibinfo
  {volume} {5}},\ \bibinfo {pages} {100} (\bibinfo {year}
  {1964}{\natexlab{b}})},\ \Eprint
  {http://arxiv.org/abs/https://pubs.aip.org/aip/jmp/article-pdf/5/1/100/19130342/100\_1\_online.pdf}
  {https://pubs.aip.org/aip/jmp/article-pdf/5/1/100/19130342/100\_1\_online.pdf}
  \BibitemShut {NoStop}%
\bibitem [{\citenamefont {Goria}\ and\ \citenamefont
  {Passarino}(2008)}]{Goria:2008ny}%
  \BibitemOpen
  \bibfield  {author} {\bibinfo {author} {\bibfnamefont {S.}~\bibnamefont
  {Goria}}\ and\ \bibinfo {author} {\bibfnamefont {G.}~\bibnamefont
  {Passarino}},\ }\href {\doibase 10.1016/j.nuclphysbps.2008.09.124} {\bibfield
   {journal} {\bibinfo  {journal} {Nucl. Phys. B Proc. Suppl.}\ }\textbf
  {\bibinfo {volume} {183}},\ \bibinfo {pages} {320} (\bibinfo {year}
  {2008})},\ \Eprint {http://arxiv.org/abs/0807.0698} {arXiv:0807.0698
  [hep-ph]} \BibitemShut {NoStop}%
\bibitem [{\citenamefont {Passarino}(2018)}]{Passarino:2018wix}%
  \BibitemOpen
  \bibfield  {author} {\bibinfo {author} {\bibfnamefont {G.}~\bibnamefont
  {Passarino}},\ }\href@noop {} {\  (\bibinfo {year} {2018})},\ \Eprint
  {http://arxiv.org/abs/1807.00503} {arXiv:1807.00503 [hep-ph]} \BibitemShut
  {NoStop}%
\bibitem [{\citenamefont {Kermanschah}(2022)}]{Kermanschah:2021wbk}%
  \BibitemOpen
  \bibfield  {author} {\bibinfo {author} {\bibfnamefont {D.}~\bibnamefont
  {Kermanschah}},\ }\href {\doibase 10.1007/JHEP01(2022)151} {\bibfield
  {journal} {\bibinfo  {journal} {JHEP}\ }\textbf {\bibinfo {volume} {01}},\
  \bibinfo {pages} {151} (\bibinfo {year} {2022})},\ \Eprint
  {http://arxiv.org/abs/2110.06869} {arXiv:2110.06869 [hep-ph]} \BibitemShut
  {NoStop}%
\end{thebibliography}%

\end{document}